\documentclass[11pt,a4paper]{amsart}

\usepackage[margin=2cm]{geometry}

\usepackage[foot]{amsaddr}
\makeatletter
\renewcommand{\email}[2][]{%
  \ifx\emails\@empty\relax\else{\g@addto@macro\emails{,\space}}\fi%
  \@ifnotempty{#1}{\g@addto@macro\emails{\textrm{(#1)}\space}}%
  \g@addto@macro\emails{#2}%
}
\makeatother

\usepackage[backend=biber, sorting=none, style=numeric]{biblatex}
\usepackage{csquotes}
\usepackage{amsmath}
\usepackage{amsfonts}
\usepackage[english]{babel}
\usepackage{graphicx} 
\usepackage{float}
\usepackage{tikz}
\usetikzlibrary{arrows, shapes}
\usepackage{xcolor}

\usepackage{array}
\newcolumntype{P}[1]{>{\centering\arraybackslash}p{#1}}
\newcolumntype{M}[1]{>{\centering\arraybackslash}m{#1}}

\usepackage{pgfplotstable}
\usepackage{booktabs}
\usepackage{array}
\usepackage{colortbl}
\usepackage{multirow}
\usepackage{adjustbox}
\usepackage{xurl}
\usepackage{framed}
\bibliography{reinsurance.bib}

\title{The barriers to sustainable risk transfer in the cyber-insurance market}
\author{Henry R.K. Skeoch$^{\ast1}$}
\address{$^1$University College London, Gower Street, London, United Kingdom}

\email{henry.skeoch.19@ucl.ac.uk}

\author{Christos Ioannidis$^{1,2}$}
\address{$^2$Aston University}
\email{c.ioannidis@ucl.ac.uk}

\address{$^{\ast}$Corresponding author}

\newtheorem{assumption}{Assumption}

\begin{document}

\maketitle

\begin{abstract}

Efficient risk transfer is an important condition for ensuring the sustainability of a market according to the established economics literature. In an inefficient market, significant financial imbalances may develop and potentially jeopardise the solvency of some market participants. The constantly evolving nature of cyber-threats and lack of public data sharing mean that the economic conditions required for quoted cyber-insurance premiums to be considered efficient are highly unlikely to be met. This paper develops Monte Carlo simulations of an artificial cyber-insurance market and compares the efficient and inefficient outcomes based on the informational setup between the market participants. The existence of diverse loss distributions is justified by the dynamic nature of cyber-threats and the absence of any reliable and centralised incident reporting. It is shown that the limited involvement of reinsurers when loss expectations are not shared leads to increased premiums and lower overall capacity. This suggests that the sustainability of the cyber-insurance market requires both better data sharing and external sources of risk tolerant capital.  

\end{abstract}

\keywords{Cyber-insurance, reinsurance, Monte Carlo simulations, efficient risk transfer, cyber-threats, security economics, insurance economics}


\section{Introduction}

Cyber-insurance has attracted considerable attention in the literature as a research topic and is now a significant insurance market in its own right, with \$6.5bn of direct written premium in 2021 in the US domestic market alone\cite{naic2021}. Commercial estimates suggest that up to 45\% of premium is ceded to reinsurers in the cyber-insurance market~\cite{sandp21}\cite{cyber2020sGallagher}. Yet the interaction between insurers and reinsurers in the cyber-insurance market has received surprisingly little attention in the literature. This paper aims to help partially address this gap by considering the asymmetry of information exchange and the uncertain time profile of damage revelation in relation to the cyber-insurance market and its interaction with reinsurers. It is then questioned whether reinsurers are sufficiently incentivised to participate in the cyber-insurance market on a long-term basis given the significant difficulties in achieving \textit{ex-post} efficient information exchange. Cyber-risk is a relatively new phenomenon and the type of attacks and their impact may change in an unanticipated manner. It is therefore important to understand the resultant issues that may arise and the ability of the market to absorb unexpected losses as otherwise the sustainability of the market is threatened.   

\subsection{Insurance market structure}
\label{subsec:market-structure}
We now briefly review the structure of the insurance market and the interaction of its various associated entities and parties. We focus in particular on capital, which is crucial for any insurance company to operate. Without appropriate capital, an insurance company cannot operate in a regulated market.

A thorough analysis of the cyber-insurance market requires the role and function of the different participants in the market to be defined. We assume here that the insurance buyer is a firm who buys insurance coverage via an insurance broker. The broker obtains quotes from different insurance firms provided by their underwriters. An underwriter is responsible for managing a book of insurance policies to deliver specified performance targets. These may vary according to the experience and skill of the underwriter (underwriters with a proven track record may be permitted to write either more premium or cover riskier entities than less experienced colleagues), the markets they cover and the risk tolerance of the provider of the insurance capital. Contrary to what might be expected, underwriting is not purely a statistical exercise. The dynamics of the exchanges between underwriters and brokers are complex, in particular with respect to information exchange which may be highly asymmetric. The job of the underwriter is to make a subjective judgement on the likelihood of the risks (prospective policyholders) they are presented with experiencing a loss and whether these can be underwritten at a premium rate which the underwriter believes is likely to be profitable. This judgement requires a certain amount of skill as while a high insurance rate is more profitable, it will attract less demand than a more attractive rate. The key objective is to price the policy such that the desired mix of risk characteristics is obtained by the insurance firm. Underwriters are assisted by actuaries, who are qualified statisticians and provide mathematical modelling services and support to assist the underwriting decisions. 

While the underwriter is the key decision maker at each insurance company in our model, insurance companies usually have multiple underwriters with different areas of expertise in terms of peril and geography - by writing policies covering different perils, insurance companies can reduce their average expected loss by diversification. Insurance brokers act as the intermediary between the insurance company and its underwriters and the end-user of the insurance. For corporate insurance, companies will typically ask their broker to prepare an insurance proposal covering a range of potential losses; these are known as lines in the insurance industry. Property, Casualty \& Professional (Liability),  Aerospace, and Maritime are well-known examples. The role of the broker is to obtain the best possible terms for its clients - both in terms of premium and depth of coverage. This requires the broker to have an excellent knowledge of the different insurance firms in the market and their reputation. Underwriters will aim to build a strong business relationship with leading brokers in the hope that they will receive a strong allocation of available premium. 

Reinsurance companies provide insurance to insurance companies. The main reason for their existence, informally, is to smooth the potential loss profile of insurance companies who otherwise might only be able to write more modest quantities of premium or hold greater capital reserves to cover potential rare outsize losses. Reinsurers also act as a potential clearinghouse for information within the market as the reinsurer will have visbility over the portfolio contents of a range of insurers (known as cedents, which rival insurance companies in the market cannot directly observe.   

Cyber-insurance presents a particularly interesting case of insurance market dynamics. The nature of the insured is particularly important as a large firm with high turnover is likely to present a more interesting and economically lucrative target for attackers, but may have better defences than a smaller firm. However, barring a systemic vulnerability the risks of significant losses in a well diversified portfolio of numerous low-limit small-medium enterprise policy may be a more profitable undertaking for a firm. An insurance company will usually obtain reinsurance to manage either tail risks associated with its portfolio (excess of loss) or to reduce its overall exposure (quota share).

\subsection{Technological advancement, information deficiency and cyber-insurance}
\begin{figure}[htb]
\centering
\includegraphics[width=14cm]{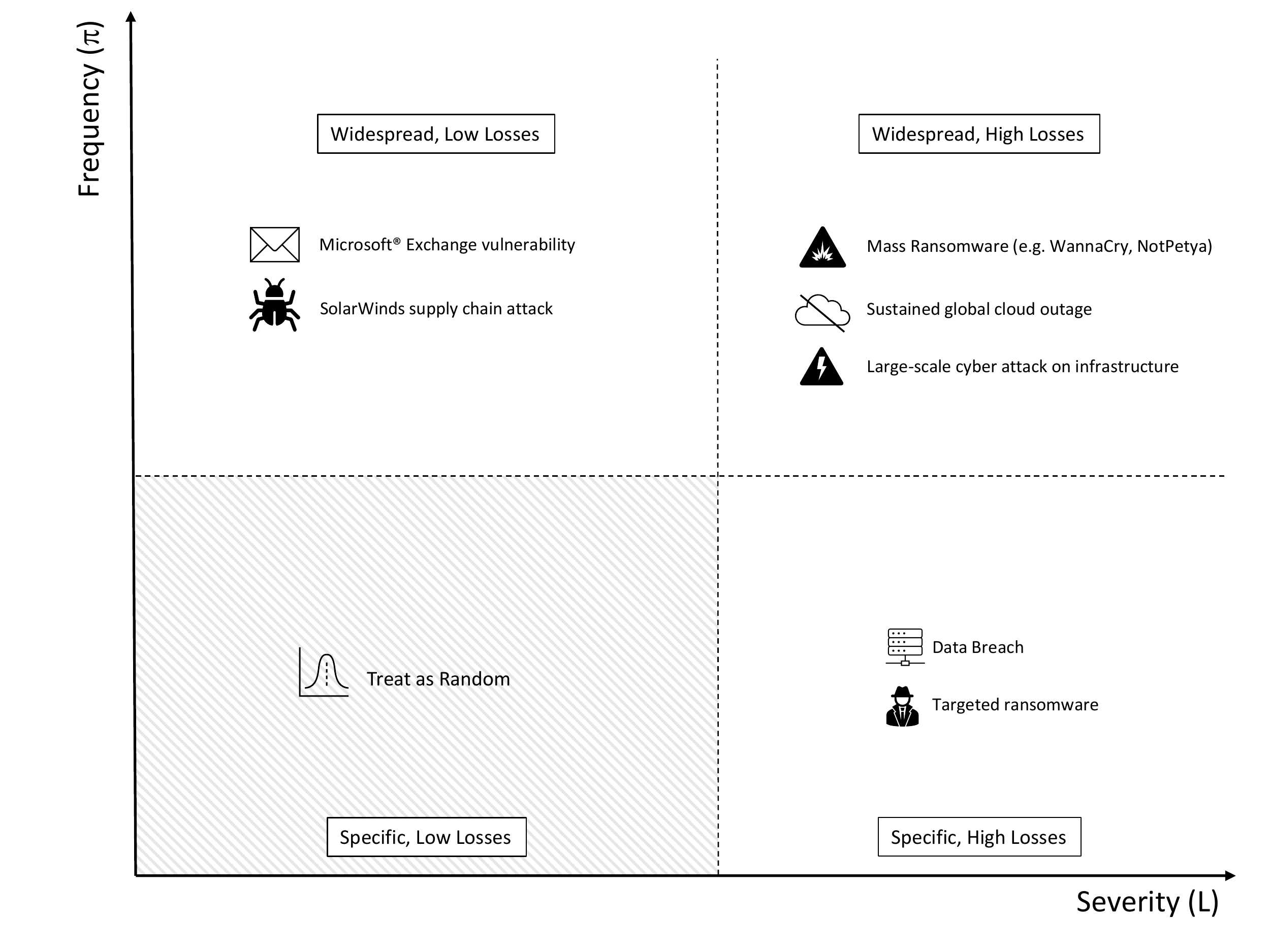}
\caption{Categorisation of cyber losses by frequency and severity}
\label{fig:attack-matrix}
\end{figure}

One particular issue for understanding loss risks stemming from cyber-attacks is the difficulty in framing the potential future scope of losses. These estimates are usually calculated by the exposure management department of an insurance company and may be either probabilistic or deterministic (based on stated realistic disaster scenarios). Exposure management traditionally is used to ascertain the risks from a natural catastrophe. In this scenario, the attacker is nature and the vectors are either wind (hurricane) or water (flooding). The questions the model for premium calculation must address are the geographical scope of the damage which determines the expected frequency of claims and the ferocity of the natural disaster which determines the expected severity. While nature is inherently unpredictable, nevertheless past experience of weather patterns gives some basis for modelling expected future losses. The relatively brief (at least in the history of insurance) history of cyber-risks and the constant evolution of technology, its integration in an ever increasing number of processes and the sophistication and capability of attackers makes such comparative predictions regarding potential losses extremely difficult. When designing cyber-insurance policies, it is important for the insurer to be highly specific in terms of the coverage and for the reinsurer to have a clear understanding of the risk dynamics it assumes if these policies are ceded. Figure \ref{fig:attack-matrix} outlines a range of possible cyber losses organised by frequency and severity. This relatively simple graphic encapsulates the potential modelling challenges associated with cyber-insurance and reinsurance. The scope for cyber-losses is determined by the evolution of technology; at the time of writing, generative artificial intelligence and quantum computing are examples of two emerging technologies that have significant security implications. 

\subsection{What claims might arise in relation to cyber-insurance?}
\label{subsec:intro-claims}

Barely a day passes without news of an emerging cyber-attack or other risk. It is important to realise that while these may be extremely disruptive for individuals, companies or societies, not every cyber-attack results in an insurance loss. An insurance loss may be defined as a loss resulting in a claim being paid by an insurer, whereas an economic loss is the total loss to an insured from the peril. Cyber-insurance policies comprise both first-party and third-party risks. A ransomware attack (without data exfiltration) would largely result in first party claims for network interruption and recovery costs. In contrast, a large data breach can incur significant third party costs. It is worth noting that such third party claims might arise several years after the policy is written. This is an important feature in cyber-insurance; a prominent relatively recent example was large hotel chain Marriott suffered a cyber-attack commencing in 2014 that was undetected until September 2018\footnote{This data breach is widely documented on the WWW from a variety of sources; for an insurance perspective see, for example, \cite{air-marriott}}, which has been one of the largest cyber-insurance claims seen thus far. A reinsurer might have expected to retain the bulk of cedent premium income after four years, only to find large claims emerging later. 

As cyber-risk is such a nascent class of business, the insurance industry is still adapting to understanding how to price the risks, which sectors are most vulnerable and how best to assess underwriting standards. This creates a risky environment to the reinsurer, particularly as cyber is likely to be a relatively small line in their overall business and they may therefore lack the requisite technical expertise to truly evaluate the risks. An interesting example is Solarwinds vulnerability\footnote{See, for example, Devanny et al (2021)\cite{devanny2021strategic}}, which proved very widespread. However, it appears that the main motivation of the attackers was espionage rather than financial gain and consequently, bar investigative costs, there is little likelihood of significant cyber-claims as a result.  

\subsection{What is the motivation for reinsurance involvement in the cyber-insurance market?}

A possible motivation for early entrants to the reinsurance market is to build market share and hope to capture premium rate increases as the market becomes more popular. As insurers gain more knowledge about the likely distribution of losses, underwriting standards may be tightened. There is nevertheless a clear information advantage possessed by the ceding insurer about the `quality' of their insurance portfolio relative to the reinsurer, which raises the issue of adverse selection. A rational reinsurer will pay extremely close attention to the information they are given by the cedent with the past loss history of the portfolio often a key feature.  The fact that so much premium is ceded suggest also that insurance carriers are themselves nervous of the quantity of risks insured relative to the likelihood of losses. This begs the question as to why reinsurers would rationally increase capital allocations to the cyber-insurance market if the originating insurer is not comfortable with the risks. One possibility is that the reinsurer may have extended scope to absorb losses from cyber-risks more readily in a diversified portfolio and may further be able to charge elevated premia if the cedent insurer is desperate to offload the risk.  

This paper aims to model the interaction between buyers of cyber-insurance, insurers and reinsurers. This is achieved via developing a model of a cyber-insurance market, which is stylized yet aims to be representative of the existing cyber-insurance market. A key argument of this paper is that there may be a diverse range of beliefs among market participants about the dynamics of cyber-risk and resultant losses. We demonstrate via simulations that under this assumption, reinsurance is only sometimes optimal for insurers. Economic theory on efficiency is consistent with this conclusion. This implies that insurers may need to rely on external sources of risk-tolerant capital (such as insurance-linked securities, or ILS). Further, there is societal benefit in better information sharing on cyber-losses, which should see convergence in beliefs. 

\subsection{Relation to existing literature}
This paper applies well-established economic theory to the cyber-insurance market in a novel manner. It is the first, to the best of our knowledge to consider the specific interaction of reinsurance capital and cyber-insurance via simulations that are representative of the existing market. 

\subsection{Paper structure}

In this paper, we examine the impact of diverse anticipations regarding cyber-losses for firms (insurance buyers), insurers and reinsurers in an artificial market. We compute the optimal involvement of reinsurance under different assumptions regarding the distribution of risks. The paper structure is as follows. In Section~\ref{sec:literature-review} we provide a brief introduction into existing literature on reinsurance, relevant actuarial models for cyber-insurance, potential sources of external and capital requirements. Section~\ref{sec:model} introduces a model for interaction between the participants in the cyber-insurance market --- buyers, insurers, and reinsurers. Section~\ref{sec:simulations} presents simulations of the artificial market under a variety of informational assumptions regarding the frequency of attacks and their severity (expected loss per attack). Finally, Section~\ref{sec:discussion} summarises our results and and presents our conclusions. 

\section{Literature Review} \label{sec:literature-review}

 This literature review commences by focusing on some useful papers for understanding the essentials of reinsurance and technical actuarial papers of direct relevance to cyber-insurance. It then considers the economic literature on efficiency, from which the core arguments of inefficiency in the cyber-insurance market are developed. 

\subsection{Reinsurance fundamentals}

Dionne et al(2013)\cite{dionne2013foundations} is an excellent collection of important papers related to reinsurance and includes many key contributions to the field. Within this, Borch(1962)\cite{borch1962} is of particular relevance to this work, focusing on describing the conditions required to achieve equilibrium in a reinsurance market via generalizing the classical theory of commodity markets to include uncertainty. Schlesinger and Doherty (1985)\cite{schlesinger1985incomplete} provide a useful treatment of issues associated with incomplete insurance markets, in particular suggesting that focusing on correlation of risks is essential for making use of incomplete markets theory. This is an argument as to why an insurer who does not currently offer cyber-insurance might enter the market should it believe that cyber-losses will not be highly correlated with areas in which it currently has exposure. Empirically, there is concern of hidden or `silent' cyber-risks within existing lines, meaning that for many insurers offering cyber-insurance could be utility detracting. Froot and O'Connell (1999)\cite{froot1999pricing} discuss the pricing of US catastrophe insurance with some illustrative data. They find that price increases and quantity declines are more pervasive than they should be within catastrophe reinsurance based on fundamental data; this is strongly suggestive of historical inefficiency. 

\subsection{Actuarial models} \label{subsec:actuarial-models}

Some interesting literature has emerged around developing specific actuarial models for cyber-insurance. Bessy-Roland et al (2021)\cite{bessy2021multivariate} introduce a multivariate Hawkes process for cyber-insurance and demonstrate how it can be calibrated using the Privacy Rights Clearinghouse database of data breaches to provide a full joint distribution of future cyber attacks (see also Hillairet et al (2021)\cite{hillairet2021expansion} for an application of such modelling to cyber-insurance derivatives). Hillairet and Lopez(2021)\cite{hillairet2021propagation} propose a stochastic diffusion model for estimating the propogation of cyber-incidents within an insurance portfolio. Biener et al (2015)\cite{biener2015insurability} outline a framework for systematically analysing the insurability of cyber-risk, concluding that there are significant hindrances towards a sustainable cyber-insurance market developing based on their criteria. Eling and Wirfs (2019)\cite{eling2019actual} use extreme value theory to estimate cyber-risk costs based on an operational risk database. In developing a model for cyber-insurance claims, catastrophic claims are a significant concern. Baldwin et al (2017)\cite{baldwin2017contagion} use the multi-variate Hawkes process as the basis of a model for estimating contagion in cyber attacks. Bessy-Roland et al (2021)\cite{bessy-roland_boumezoued_hillairet_2021} introduce a multi-variate Hawkes framework for modelling and predicting cyber attacks frequency across firms following successful cyber-attacks against a subset of the population.   

\subsection{Economic Theory on Efficiency}
\label{sec:econefficiency}
We now consider how to relate well-established economic arguments on efficiency to insurance of cyber-risks. A rational buyer of insurance will likely aim to purchase a policy via a market in order to achieve a price they deem acceptable (ideally optimal). The aim of a well functioning market is to match buyers and sellers of a particular good and to establish a fair price for that good. Efficiency is often used as a measure for the efficacy of risk transfer in a market and can be defined in two ways: \emph{ex-ante} (before the transaction) or \emph{ex-post} (after the transaction). Ex-ante efficiency requires conditions, that we shall demonstrate are extremely hard to satisfy. Ex-post efficiency can be realised but requires an exchange of information. It should be noted that an efficient ex-post premium if realised would be considered an actuarially fair premium; this is distinct and different from a premium that meets customer expectations and is subjectively viewed as acceptable based on risk tolerance or beliefs. A lack of efficiency does not mean that transactions will not take place, but creates a comparative advantage for the party with greater access to, or possession of less noisy, information. 

\subsubsection{Ex-ante efficiency}

According to the Arrow-Debreu model\cite{Arrow1965}\cite{debreu1959theory}, a complete market has:

\begin{enumerate}
	\item Negligible transaction costs and therefore perfect information
	\item Every asset has a price in every possible state of the world
\end{enumerate}

Both of these assumptions are highly unlikely to be valid for cyber-insurance markets. For an asset such as a stock or bond, which may be continually traded, price is a legitimate marker of information. However, commercial insurance contracts are struck at discrete time periods and are valid for a specific length of time only. These typically operate on a yearly basis with key renewal points throughout the year dictated by market convention. Further, there is a significant cost of operating for the insurance company that is typically passed onto the customer via the premium. Most insurance contracts are non-fungible and non-transferable, unlike many publicly traded financial instruments such as stocks or bonds. This is a fundamental feature of insurance markets and implies that the first condition of completeness within the Arrow-Debreu model is unlikely to be satisfied. The second assumption that every asset has a price in every possible state of the world is equally not realistic as in reality insurance companies may decline to quote for a particular policy if the insuring party considers the risks outside of their tolerance. 

\subsubsection{Ex-post efficiency}
\label{subsubsec:ex-post-efficiency}
Starr (1973)\cite{starr1973optimal} suggests that a set of valuation decisions is ex-post efficient if that “there be no redistribution that will increase some trader’s realized utility while decreasing no trader’s realized utility”. Alternatively, as interpreted by Feiger (1976)\cite{feiger1976diverse}, “there exists no alternative feasible set which is sure to be Pareto improving, looking back from the state which actually occurs.” The Arrow interpretation of states of the world is convenient for an insurance analysis as certain states of the world are loss-triggering. There are a diverse range of possibilities for attempting to frame these states – one possible utility driven approach is to model the utility of the protector of an information set using confidentiality, integrity and availability and constructing potential attacks degrading these properties in terms of deviations from their preferred state. A cyber-insurance policy can cover a wide range of potential losses, an interesting case being costs of specialist IT consultants to help diagnosis and recovery after a data breach, for example. A data breach is an attack primarily on confidentiality but if the system from which the data is taken is somehow modified by a malicious actor to facilitate the theft, then it is also an attack on integrity. Recently, ransomware attacks have become a prominent cyber-threat adding a further risk of loss of availability. 

A particular issue for cyber-insurance is the risk of a catastrophic cyber event. A problem with establishing distributions for catastrophic events is that the sample space is often sparse as these events tend not to occur too often. Despite computer systems and networks being societally ubiquitous in most developed countries, public data about cyber-attacks and computer mishaps of the standard required to properly price cyber-insurance contracts remains lacking. Returning to the definitions of Starr and Feiger, these require careful interpretation in the context of cyber-insurance. Consider the scenario in which an entity suffers a loss as a result of a cyberattack, which is deemed `with high confidence' by relevant National Cybersecurity Agencies to have been state sponsored. In an efficient market, it ought to be the case that a loss is experienced and thus constitutes a valid claim. However, many insurance policies contain what is known as a `war clause'. This is a relatively obvious protection in the case of conventional property damage policies but is harder to interpret in the context of cyber-insurance and state sanctioned if not directed cyber-operations. In the event of a significant cyber-attack, the world reaches a state whereby losses are generated. These claims ought to be paid, yet there is a clear issue in terms of potential legal action - as seen in recent court cases involving Merck and Mondelez where both parties suffered multi-billion dollar economic damages as a result of malware believed to have originated from nation state-backed entities. Consequently, there exists a clear path towards violation of both the Starr and Feiger conditions. In sum, when there is not certainty about a claim validity, ex-post efficiency is de facto unachievable. 

\subsubsection{Rational belief equilibria}

Kurz (1994)\cite{kurz1994rational} compares rational expectations equilibria, in which all agents know the true probability distribution of prices, with rational belief equilibria, in which no one knows the true distribution of prices and each agent must form their own belief about it. Even at first sight, it appears intuitive that the latter category of equilibrium is likely to better characterise cyber-insurance decisions given that a claim to know the path of future technological development with even a degree of confidence is almost certainly fallacious. Kurz's theory of rational belief equilibria relies on the system being stationary for the purposes of agents generating forecasts. The theory identifies a set $B(Q)$ of beliefs compatible with the data generated under $Q$, which cannot be rejected by the data. At first sight, this may appear a significant issue for analysis of cyber-risk. However, one possibility is that there exists a brittle equilibrium for a finite period of time, subject to shocks. Eventually a shock, or paradigm shift in the sense of Kuhn(1962)\cite{kuhn}, may perturb the market from its state of equilibrium. This causes market participants to abandon their beliefs but then upon stabilisation a new set of beliefs may be formed. For example, the ransomware epidemic post-WannaCry makes for an interesting case study. This introduced a hitherto less well considered generator of potential losses, which insurers had to adjust for in their policies and subsequently triggered a marked increase in premiums charged to the market.  

\section{Model} \label{sec:model}
We now introduce a model for describing the dynamics of a reinsurance market. We use standard results in the microeconomic theory of insurance without derivation for brevity. The motivation for this is to outline in formal economic terms the structure of an insurance market with reinsurance, from which theoretical simulations may be developed.  
 \subsection{Insurance buyer} \label{subsec:insurance-buyer}
 Before formulating the model for a market, we establish the baseline decision of a buyer of insurance facing two states --- loss and no loss --- with probability $\pi$ and $1-\pi$, respectively. The corresponding utility function is
\begin{equation}
E[U] = (1-\pi)u(W-P(C)) + \pi u(W-P(C)-L+C-D)
\label{eq:baseline-buyer-utility}
\end{equation}
$u$ is the constant absolute risk aversion (CARA) utility function, 
\begin{equation}
    u(w) = \frac{1 - e^{-\alpha w}}{\alpha} 
\end{equation}
where $\alpha$ is a constant. For the purposes of this research, we chose CARA as it is a commonly used utility function and sufficiently captures the trade-offs we wish to model. Other forms of the utility function might be deployed to represent more complex buyer preferences. The parameters in Equation~\ref{eq:baseline-buyer-utility} are $W$, representing the initial wealth of the insurance buyer; $P(C)$, the premium paid for an amount of insurance coverage, $C$; and $D$, the deductible\footnote{The amount of losses which must be borne by the insurance buyer} set by the insurer. We shall assume that
\begin{equation}
P(C) = pC
\end{equation}
where $p$ represents a premium \emph{rate}. We emphasise that the customer chooses the coverage amount $C$, up to a limit permitted by the insurer and observes the premium rate, $p$, from different insurance companies. $L$ is the loss experienced in the loss state. In the event that there are multiple loss states, denoted by $s$, we assume that these belong to a finite and countable set of states, $S$, such that $s \in S$, with a corresponding loss for that state, $L_s$. Specifying an initial endowment, $W_0$, and representing the total cash premium paid as $P$, Equation~\ref{eq:baseline-buyer-utility} may be restated

\begin{equation}
E[U] = (1 - \sum_s \pi_s)u(W_0-P) + \sum_s \pi_s u(W_0 - P - L_s + C_s - D_s)
\label{eq:baseline-buyer-utility-state-dep}
\end{equation}
Both Equations~\ref{eq:baseline-buyer-utility} and~\ref{eq:baseline-buyer-utility-state-dep} are equivalent and for the unsophisticated cyber-insurance buyer, Equation~\ref{eq:baseline-buyer-utility} is a sufficient formluation of the problem. However, when considering the supply dynamics of the cyber-insurance and reinsurance markets, it would be expected that the insurance company consider the different states that may be loss generating. We assume that the objective of the insurance buyer is to maximise their utility.
\begin{framed}
\begin{assumption}
    The insurance buyer aims to maximize their utility
\end{assumption}
\end{framed}

\subsection{Supply of insurance} \label{subsec:insurance-supply}
Having established the theoretical decision framework for the insurance buyer, we now establish a formal model determining the supply of cyber-insurance. Following Hammond (1981)\cite{hammond1981ex}, we consider the actions of consumers in the economy:
\begin{equation}
	x^i(s) = [x^{i}_t, x^{i}_{t+1}(s)] 
\end{equation}
$i$ represents an individual consumer of a total $I$ consumers in the marketplace. As before, $s$ represents a contingent state of the world, and it is assumed that the set of possible states, $S$ is finite. The vector of total insurance demand, $\mathbf{x}_t = [C^1_t,C^2_t,\dots,C^i_t]$. 

We assume that there are $J$ insurers in the market, each with a supply of insurance
\begin{equation}
	\mathbf{y}^{j}(s, \mathbf{x}) = [\mathbf{y}^{j}_t(\mathbf{x}_t), \mathbf{y}^{j}_{t+1}(s, \mathbf{x}_{t+1})] 
\end{equation}
$\mathbf{y}^j_t$ is an i-length vector of the units of insurance sold by insurer $j$ to customer $i$ at time $t$ and consequently, which, expressed in monetary terms is identical to cover, $C$. It is assumed that each customer $i$ has an exclusive policy with its chosen insurer $j$. 
Each insurer has a premium vector,
\begin{equation}
    \boldsymbol{\pi}^j = [p_1, p_2, \dots, p_i]
\label{eq:premium-vector}
\end{equation}
representing the premium rate it charges to each customer. This vector may be time dependent. For conciseness of presentation, we will henceforth drop time subscripts as the analysis in this paper is restricted to a single period.  

\subsubsection{Insurer objectives}

We assume the insurer formulates its decisions on insurance supply, $\mathbf{y}^{j}(s, \mathbf{x})$ via the following parameters (see Chapter 3.5 of\cite{rees2008microeconomics}):

\begin{itemize}
    \item $K$: the reserve capital held by each insurer.
    \item ${\pi}$: the total premium income for each insurer.
    \item $X$: the claim costs (losses) experienced, described by probability function $F(X)$ with differentiable density $f(X)$ defined over the interval $[0, X_{\max}]$. 
    \item $D$: the total deductible enforced by the insurer.
    \item $W_0$: the initial wealth of the insurer - this may be thought of as shareholder equity, for example, or syndicate (non-regulatory) capital.
    \item $W$: the residual wealth the insurer has after paying claims. If the amount of claims is greater than $A \equiv \pi + K + W_0$, the insurer faces ruin. 
    \item $r$: the risk-free interest rate for the relevant period.
\end{itemize}
We assume that each insurer has zero utility condition and its objective is to maximise its wealth
\begin{equation}
    W_i = W_0 + \pi_i + D_i - \int_0^A{\frac{C_i}{r}}dF_i(X)
\label{eq:insurer-wealth}
\end{equation}
subject to the constraint
\begin{equation}
    W_i + K_0 > 0
    \label{eq:ins-capital-constraint}
\end{equation}
The optimal set of allocations for the insurer would be to policies that maximise the wealth/capital ratio $W_i/K$. 
 \begin{framed}
     \begin{assumption}
         Insurers aim to maximise their wealth
     \end{assumption}
     \begin{assumption}
        \label{assumption:de-finetti}
         The probability distribution, $F_i(X)$ is subjective to each insurer in the sense of de Finetti (1974)~\cite{definetti1974}. This will be justified in Section~\ref{subsec:modelling-cyber-risks}.
     \end{assumption}
 \end{framed}

\subsection{Introducing reinsurance} \label{subsec:reinsurance-model}

In order to reduce risk exposure, the insurer may also seek to purchase reinsurance. There are two categories of reinsurance considered in this work: quota share and excess-of-loss. Reinsurers are consequently concerned with determining the probability of two types of extreme events: those resulting in single large losses from a particular client (concentrated losses) and those resulting in widespread repeated claims across cedents (contagion). In the event that this distribution is objective, then this would lead to a universal fair price for insurance. Reinsurers in turn will have their own subjective distributions and charge the expected value of their own distributions to clients. While this may be commercially reasonable, such prices are not fair in a strict economic sense. The existence of reinsurance serves to allow insurers to smooth their subjective expected loss distributions, which clearly implies risk aversion as opposed to neutrality. In short, intermediation implies imperfection\footnote{Skiadas (2013)~\cite{skiadas2013smooth} presents an interesting analysis on this topic}. Including reinsurance, Equation~\ref{eq:insurer-wealth} becomes:
	\begin{equation}
W_i = W_0 + (\pi_i - R_i) + D_i - \int_0^A{\frac{C_i-I_i}{r}}dF_i(X)
\label{eq:insurance-supply-withre}
\end{equation}
where the parameters are as above, with the addition of $R$, which represents the cost of reinsurance to the insurer and $I_i$, which is the amount of losses indemnified by the reinsurance policy purchased. The constraint $W_i + K > 0$ continues to apply. Notation-wise, in similar fashion to Section~\ref{subsec:insurance-supply}, we use vectors to describe reinsurance supply. We assume that there are $k$ reinsurers, who charge $\mathbf{r}^k$ rates to insurer $j$ and denote the supply vector of reinsurance as $\mathbf{z}^k$.

For a simple quota share policy, \[ R = \rho P \] where $\rho$ is the proportion of the portfolio ceded and then \[ I(L) = \rho L \]
However, in cases involving excess of loss or other reinsurance treaties, the calculation is more involved. Miccolis (1977) \cite{miccolis1977theory} provides an exposition of some standard mathematical techniques for describing excess of loss calculations. In the case of excess of loss, the indemnification equation becomes: 
\begin{equation}
	I(L) = (L - B)^+ - (L - B - A)^+
\end{equation}
$A$ and $B$ are parameters for an excess of loss policy covering \$A(mn) of losses in excess of \$B(mn). For simplicity, it is assumed that each insurer can purchase only a single excess-of-loss policy from each reinsurer. It would seem rational for the purposes of our discussion that the insurers seek to buy reinsurance above the aforementioned value $A$, losses above which the firm becomes insolvent. 

\subsubsection{The reinsurance market} \label{subsec:reinsurance-market}

We assume that there are $k$ reinsurers in the market who provide reinsurance capacity. The reinsurer aims to maximise wealth in similar fashion to the insurer (Equation~\ref{eq:insurer-wealth}), but does not include a deductible:

\begin{equation}
W_r = W_0 + \pi_r - \int{\frac{I_r(X)}{r}}dF_r(X)
\label{eq:reinsurance-supply2}
\end{equation}
$\pi_r$ is the total reinsurance premia received and $I_r(X)$ denotes expected reimbursements paid out to cedents. The reinsurer is subject to the capital constraint $W_r + K_r > 0$. Note that we allow for the reinsurer and insurer to have different beliefs about the expected distribution of losses.

\begin{framed}
    \begin{assumption}
        The reinsurer may have a different belief from the insurer regarding the distribution of risks. 
    \end{assumption}
\end{framed}

\subsection{Modelling cyber-risks}
\label{subsec:modelling-cyber-risks}
We have thus far considered losses related to cyber-risk in an abstract sense as setting up the theoretical framework for evaluating the interaction between buyers, insurers, and reinsurers does not require the functions dictating these losses to be instantiated. However, simulating the decision making to analyse the potential for efficiency in the market does require some sample distributions. We use standard results in probability theory without derivation (the reader wishing to understand the background more thoroughly is referred to any standard statistical text on probability theory; Williams (1991)~\cite{williams1991probability} is a particularly accessible and carefully explained introduction). While using formal probability theory is not essential for simulating the results in this paper, it is beneficial to apply theoretical rigour as this helps to highlight some features specific to cyber-risk that are potentially problematic for formulating traditional actuarial insurance assessments.  

We start by defining a probability triple $(\Omega, \mathcal{F}, \mathbf{P})$. $\Omega$ is a set representing the sample space of \emph{all events}. $\omega$ represents a sample point of the sample space. The $\sigma$-algebra\footnote{The definition of a $\sigma$-algebra is a collection of subsets of a set that is closed (stable) under any countable number of set operations. This is important for working with probabilities, where the probabilities of all possible outcomes must sum to 1.}, $\mathcal{F}$, on $\Omega$, is known as the family of events\footnote{See Chapter 2 of Williams (1991)~\cite{williams1991probability}}. Denoting an event by $A$, we may write
\begin{equation}
    \mathcal{F} = \{A | A \subseteq \Omega, A \in \mathcal{F} \}
\end{equation}
The intuitive explanation in relation to cyber-insurance is that $\mathcal{F}$ is the collection of events covered by a policy that may trigger a claim and then, possibly, a loss to the insurer. If $\mathcal{F}$ is the Borel\footnote{The Borel $\sigma$-algebra, $\mathcal{B}(\mathbb{R})$, is the smallest $\sigma$-algebra containing all open intervals in $\mathbb{R}$} $\sigma$ algebra on the set of real numbers, then there exists a unique probability measure on $\mathcal{F}$ for any cumulative distribution function. Letting $X$ be a random variable on $(\Omega, \mathcal{F}, \mathbf{P})$, 

\begin{equation}
    \begin{aligned}
         \Omega \xrightarrow{X} & \mathbb{R} \\
         [0,1] \xleftarrow{\mathbf{P}} \mathcal{F} \xleftarrow{X^{-1}} & \mathcal{B} \\
    \end{aligned}
\end{equation}
Informally, this means that so long as there is a collection of events that obeys certain mathematical properties, it is possible to assign a probability to an event using a probability distribution function. One interesting outcome is that a key assumption of probability theory is that the system is stable. This is a potentially problematic assumption for cyber-risk as there have been clear examples of previously unconsidered threats developing. However, insurance policies comprise a set of event definitions as part of the policy, which are contractually binding (albeit open to legal dispute). The importance of careful policy wording is consequently readily apparent. As will shortly be explained, underwriting cyber-insurance policies requires an assumption of subjective, temporary stationarity in distributions. This is a realistic assumption in the context of industry practice, where (re)insurance policies last for a year and then are re-priced based on updated distributions resulting from supply-demand dynamics and claims experienced.  

\subsubsection{Why use subjective probabilities to model cyber-risks}
Assumption~\ref{assumption:de-finetti} in Section~\ref{subsec:insurance-supply} stated that the probability distributions that govern insurance supply are subjective in the sense of de Finetti (1974)~\cite{definetti1974}. We now provide the intuition behind and justification for this assumption before moving to consider the form of distribution that might be used to model cyber-insurance decisions. 

In Section~\ref{subsubsec:ex-post-efficiency}, we outlined the conditions required for ex-post efficiency. Considering these in the context of cyber-insurance, we conclude that ex-post efficiency is unlikely to hold and almost certainly cannot be implemented at the time when the underwriting decision is made. Unless of course, the true probability distribution attached to the known and finite states of nature is known and shared by all participants. Such condition is the foundation of the theory of rational expectations.  This is synonymous with the existence of a stationary distribution. One way of defining a stationary process is to say that its moments are time-independent, which means that the average value of the measurements is a constant.  Such distributions are foundational for the existence of efficient equilibria under risk.

It is usual in macroeconomics to depict technological progress as a Markov chain. If the depicted process has started far from its invariant distribution, then it is also non-stationary, but easy to predict as it will approach the limiting distribution that is ultimately stationary. However, in a short epoch, it will appear as non-stationary. Whether technological progress has such a limiting distribution is an unresolved question. Over the long-run it appears to have exhibited a definite trend, with some downwards transitions attributable to natural disasters, wars epidemics etc.. In the short run, local approximations can be derived, and expectations can be formed, however agents will splice different segments depending upon their horizons and discount rates. The imposition of rational expectations restrictions upon this structure can only be justified if all agents have identical preferences and endowments, a condition that by construction does not hold. For Markov chains with non-stationary transition probabilities, no steady-state typically exists and almost nothing in the non-stationary setting is computable in closed-form. 

It is hard to imagine that there is any way to truly predict an arbitrary non-stationary process. This is because as soon as one postulates a future path another can always reverse it, without creating any problems of consistency with earlier data. In a more general case one might lower expectations, not to actually predicting well, but to predicting with low regret. To this effect agents can choose their most suitable approximations selecting the time span and use their best computational algorithms.

In the absence of a universally accepted probability distribution, \textit{ex-post} efficiency is almost impossible to attain.  Of course, there are opportunities which can best utilised only with \textit{ex-ante} knowledge of the state of nature. In its simplest form, it is the choice of technique in production/product that depends upon the expected state. However, a more interesting situation arises when the expected state conditions the preferred level of production. Expecting the cyber-insurance market to quote premia at all levels that are consistent with ex-post efficiency is rather unrealistic. The very nature of the underlying processes does favour the existence of a generally accepted stationary distribution. Rational agents will behave as if they are \textit{ex-ante} efficient using their own expectations of losses based on their subjective probability distributions taken over their own sample spaces. 

The evolution of cyber-threats will be conditioned of the path of technological improvements in both elements of information and communications technology, software and hardware.  The future path of such advances may be partly predictable based on well-established empirical regularities, such as Moore’s law that famously predicted that the number of transistors on integrated circuits would double every two years, i.e. at an annual rate of about 40\%~\cite{moore2006moore}. Others\footnote{For example, Benson and Magee (2015)~\cite{benson2015quantitative}, Funk and Magee (2015)~\cite{funk2015rapid}, and Nagy et al (2013)~\cite{nagy2013statistical}}, looking at related data came to the conclusion that predictions of particular technological IT innovations, such as hard drives may be approximated using exponential functions.  A very useful exposition of this attempt, using smooth functions the predict technological progress is Farmer and Lafond (2016)~\cite{farmer2016predictable}.

Yet technological advances undergo structural breaks, where both the level of technology in terms of some of its main characteristics and its future direction change. A prominent example at present is the introduction of quantum computing, which will alter radically reduce computational time and thus has implications for the robustness of cryptographic protocols that are currently infeasible to attack on a realistic timeframe.

Technological progress is achieved by the complex interactions of two main human pursuits. The organised knowledge as it appears in scientific papers, submitted patents, recipes, protocols, routines and probably informal know-how, acquired through `learning by doing' in a long process of imitation and repetition. The development of science, technology, innovation and production require both codification and knowledge. 

It seems unlikely that such dual processes can be tamed into a smooth parametric function with time invariant parameters, shared by all participants. If anything, in the absence of such shared beliefs, it is expected that for market participants whose welfare depends upon such developments, their decision making will be based on arbitrarily diverse anticipations. These are individually efficient decision makers because they act on the basis of all the information available at the time. It is clear therefore that by and large insurance contracts on expected losses based of future technological developments, that are subject to structural changes, cannot be written on generally accepted parameters, to deliver Arrow-Debreu type ex-ante efficient premia. All the participants are efficient in terms of fully exploiting their private anticipations of losses, but the quoted premia at the two levels will not result  fully efficient in the Pareto sense economic outcomes.

\subsubsection{Probability Distributions}
For the simulations in Section~\ref{sec:simulations}, we separate the expected distribution of losses into the number of expected claims (frequency) and the average expected loss per claim (severity). This is a very common method for actuarial modeling and is described in most standard texts, for example Panjer (2006)~\cite{panjer2006operational}. Its appropriateness to categorising cyber-risks was described in Section~\ref{subsec:intro-claims} and summarised in Figure~\ref{fig:attack-matrix}. We assume that frequencies follow a Poisson distribution and severities a log-normal distribution. The Poisson distribution is a standard starting point for frequency modelling (see, among many, \cite{MikoschThomas2004Nim:}). There is no clear consensus in the empirical literature on which distribution is most appropriate for describing the severity of cyber-losses (see in particular \cite{eling2019actual, woods2021county}). We use the log-normal distribution as a starting point as it is well-understood and straightforward to configure. We use simulated rather than empirical distributions as the aim of the simulations is to examine whether efficiency is theoretically possible, whereas markets in practice are very unlikely to be efficient. The probability of $k$ events occurring in a unit of time represented by the Poisson distribution is
\begin{equation}
    f(k,\lambda) = \frac{\lambda^ke^{-\lambda}}{k!}
\end{equation}
where $\lambda$ is the expected number of events. The log-normal distribution assumes
\begin{equation}
    \ln(X) \sim \mathcal{N}(\mu, \sigma)
\end{equation}
that is the natural logarithm of variable $X$ is normally distributed with mean, $\mu$, and standard deviation, $\sigma$, which are defined as
\begin{equation}
    \mu = \frac{\mu^2_X}{\sqrt{\mu^2_X+\sigma^2_X}}\quad \text{and} \quad \sigma^2 = \ln\left({1+\frac{\sigma^2_X}{\mu^2_X}}\right) 
\end{equation}
$\mu_X$ and $\sigma^2_X$ are the mean and variance, respectively, of the variable $X$. The probability density functions and cumulative distributions functions for the log-normal distribution are readily available in any standard resource on statistics and are omitted for brevity. 

\subsubsection{Combining probability distributions}
Cyber-insurance policies cover a diverse range of first- and third-party risks and consequently, there is probably no one distribution that actually covers all relevant risks. Accordingly, it is desirable to consider a combination of possible risks. Unfortunately, probability distribution functions are rather difficult to combine with a closed-form solution (see, for example, Nadarajah et al (2018)~\cite{nadarajah2018sums}) and require analytical solutions. A common strategy is to use a package such as Mathematica~\cite{wolfram1991mathematica}. However, there is an alternative approach which is to use Monte Carlo-type simulations. Section~\ref{sec:simulations} will show how these can be deployed to yield useful insights on insurance decisions, the results of which do not require sophisticated mathematics to formulate or interpret. 

\section{Simulations}
\label{sec:simulations}
We consider simulations of a cyber-insurance market with reinsurance over a single period. We assume that losses arise in the period of the insurance policy and are recorded at the time they arise. Policy data is confidential to insurance companies and consequently, the simulations are established for model convenience but are constructed to replicate real-world insurance market dynamics. We use Poisson distributions for the frequency of losses and log-normal distributions for the severity of losses (details of these distributions and their associated functions may be found in any standard statistical text). The Poisson distribution is a common choice for modelling claim frequencies in insurance (see, among many excellent references, \cite{panjer1992}\cite{MikoschThomas2004Nim:}). There is no clear consensus in the literature on the optimal distribution for modelling the severity of cyber-related claims, but the log-normal distribution has been shown to be a reasonable approximation in the limited empirical studies to date (e.g. Eling et al (2019)~\cite{eling2019actual}, Woods et al~(2021)\cite{woods2021county}. The use of the joint frequency-severity distribution approach follows Panjer (2006)~\cite{panjer2006operational}. We assume a common set of contracts across insurers varying in limit size.     

The analysis considers only variation in coverage and premium. We assume arbitrarily a market size of \$500mn total coverage. The simulations were computed using the Julia programming language. We found the \textit{Distributions.jl}\cite{Distributions.jl-2019}, \textit{QuadGK}\cite{quadgk}, and \textit{Plots.jl}\cite{PlotsJL} packages particularly useful in facilitating the presentation analysis. Unless otherwise specified, \textit{Monte Carlo} type simulations were run 100,000 times.  

The goal of the simulations is to illustrate how capital supply from the reinsurance market to the insurance market and then to buyers is inherently inefficient as pricing is influenced by the diversity of opinions regarding the frequency and severity of losses even with relatively simple standard distributions. The simulations might be applied to a variety of insurance markets, but they have been constructed to be representative of the existing cyber-insurance market. The authors subjectively note based on their interactions with insurance market professionals that existing catastrophe models, for example, are being referenced as a starting point for considering extreme cyber-risk scenarios. We believe this modelling represents a useful contribution as there is currently significant divergence in commercial cyber-industry model estimates according to a 2023 report by a leading insurance broker, Guy Carpenter~\cite{guycarp2023}.

\subsection{Preliminaries}
Familiarity with the insurance market is not a prerequisite for understanding and interpreting the simulations that follow. We have taken care to explain the terms used and ensure parameters are fully defined and explained. However, the reader unfamiliar with corporate insurance may find the following definitions helpful as a reference. These may be safely skipped for those experienced in either the practice or study of insurance. 

\begin{itemize}
    \item $\mu_L$: The average expected loss in monetary (cash) terms.
    \item $\sigma_L$: The standard deviation of losses in monetary (cash terms).
    \item $F^{-1}(p)$: The loss value that occurs with probability $p$ according to the cumulative distribution function $F$. If $p=0.95$, then in 95\% of cases, the loss is expected to be less than or equal to the output of this function. 
    \item Loss ratio: the percentage of cash premiums collected by an insurance company for a specified period (usually a year) paid out as losses.
    \item Frequency: the number of claims in a period.
    \item Severity: the average loss per claim.
    \item Cover/Exposure: the total maximum losses that could result from a policy/portfolio respectively.
    \item Expected loss: the mean loss from a policy/portfolio.
    \item Technical premium: the cash premium or premium rate (calculated as the ratio Expected Loss/Cover) corresponding to the expected loss. This is the premium income at which the insurer can be expected to break even. 
    \item Simulated loss: the average loss from running $N$ simulations based on random sampling of the expected loss distribution. This can only be computed once the portfolio is formed, so we assume that premiums are calculated based on expected loss values. 
    \item Ceding commission: the percentage premium paid back to a cedent by a reinsurer to cover underwriting expenses and other costs. 
\end{itemize}

It is important to note the sequencing of the insurance transactions in the market. The insurance buyers observe a premium rate and based on this decide how much cover to buy. The insurance provider then has obtained a portfolio. Based on the risk characteristics of that portfolio, the insurer may look to enter into a reinsurance contract to eliminate some potential risk. The simulations assume that insurers and reinsurers target a specific loss ratio \textit{ex ante} to determine pricing.

\subsection{Simulation Strategy}
We consider three simulations:

\begin{enumerate}
    \item A benchmark simulation.
    \item One reinsurer, five insurers with different portfolios comprised of different weights of five common contracts, buyers not considered.
    \item One insurer, one reinsurer, different buyer price sensitivities. 
\end{enumerate}

These simulations are distinct from each other, though have broadly consistent parameters where possible. The aim of the benchmark simulation is to demonstrate the approach used to generate loss distributions and also to instantiate buyer utility functions to show that if the buyer has a different expectation of loss severity from the insurer, then full insurance coverage may not be utility maximising. 

The second simulation starts with a reinsurer who has a range of distributions its actuaries consider acceptable. The reinsurer attempts to offer reinsurance to achieve a target loss ratio and so quotes a reinsurance rate to the market. The market consists of five insurers who have portfolios that range from a large number of small loss risks (called Insurer Alpha) to a small number of large loss risks (called Insurer Echo) with Insurers Beta, Charlie and Delta having portfolios that move progressively between the two extremes. This aims to replicate the structure of the cyber-insurance market in a stylised form and contrast the appropriate reinsurance strategy for the different types of insurer. 

It should be noted that the premia in the simulations may vary from those witnessed in the market and in some cases appear very large. The simulations are intended to guide the reader through an application of the economic theory and market model from a theoretical perspective and demonstrate the difficulty of establishing efficiency rather than aiming to be a simulation of the real-world cyber-insurance market. 
\subsection{Simulation 1: Benchmark simulation}
\begin{figure}[htb]
\centering
\includegraphics[width=16cm]{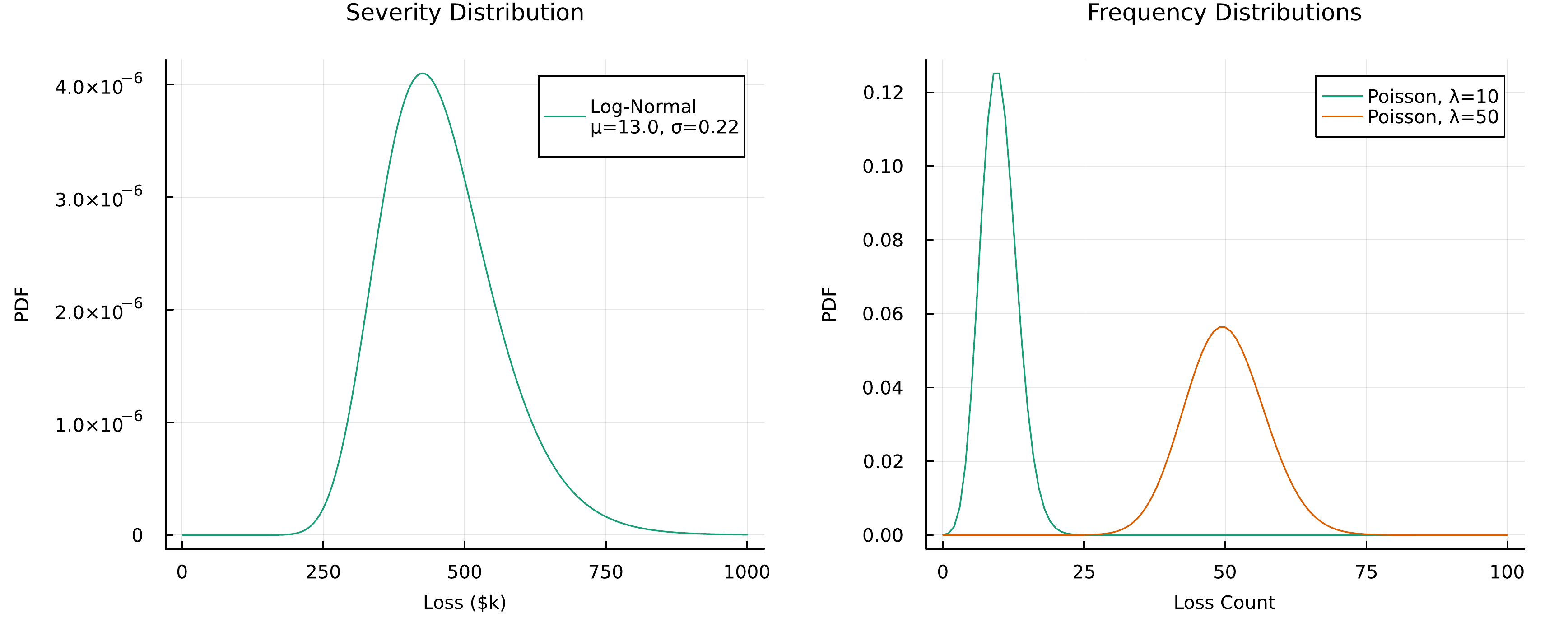}
\caption{Benchmark severity (LHS) and frequency (RHS) distributions}
\label{fig:bmk-sev-freq}
\end{figure}

\begin{figure}[htb]
\centering
\includegraphics[width=12cm]{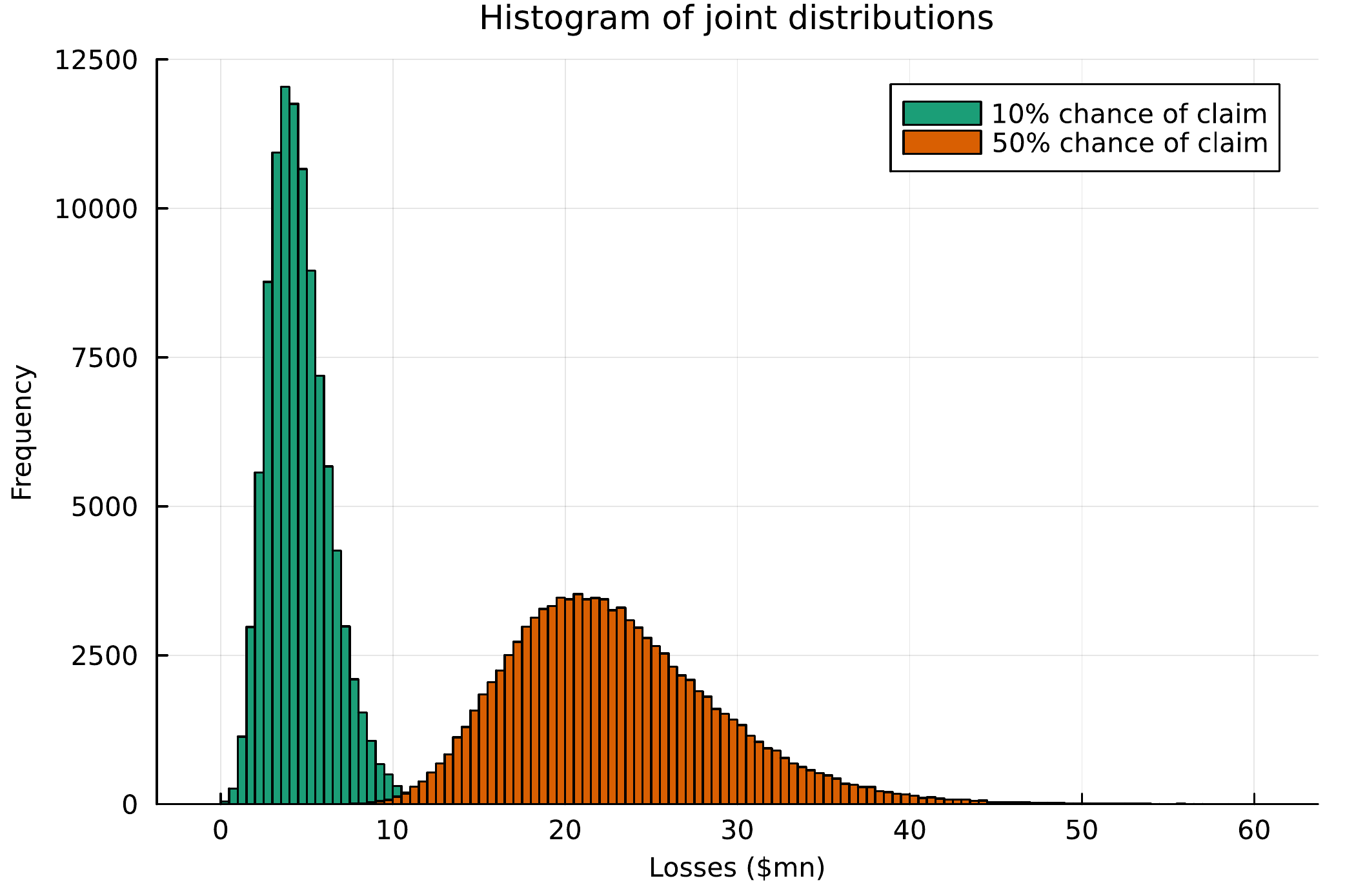}
\caption{Simulated loss distributions}
\label{fig:bmk-joint-distro}
\end{figure}

We first consider a simple simulation before starting to examine the effects of varying market structure and pricing variables. This simulation assumes the following:

\begin{itemize}
    \item There is only one insurance policy offered in the market, with a limit of \$1mn.
    \item The mean expected loss (severity) for each policy is \$500k, with standard deviation \$250k.
    \item We consider two scenarios --- one where 10\% of policies are expected to experience a loss and another where 50\% of policies are expected to experience a loss.
    \item There are 100 buyers, five insurers and one reinsurer in the market. For simplicity, we model total losses for the market and assume they are evenly distributed. 
    \item Losses are simulated with 100,000 runs and random sampling of the severity and frequency distributions.
    \item Distributions are shared by all market participants.
\end{itemize}
Figure~\ref{fig:bmk-sev-freq} plots the probability distribution functions of the severity distribution and the two frequency distributions. The severity distribution is log-normal with parameters $\mu=13.0$ and $\sigma=0.22$; the two frequency distributions are Poisson with $\lambda$ of 10 and 50, respectively. The PDF values for the severity distribution are very small because of the units of the loss; the integral of the PDF across the function domain must sum to 1. Running a simulation, the expected loss distribution for the two frequency distributions can be obtained. This is presented in Figure~\ref{fig:bmk-joint-distro}. The values on the y-axis of Figure~\ref{fig:bmk-joint-distro} simply represent the number of times each loss value range in the histogram appears in the simulation. Each bar in the histogram has a width of \$0.5mn. This is simply chosen for aesthetic reasons. The main emphasis is on the shape of the distributions rather than the precise frequency count in the histogram. 
 
Having examined the distributions, we now consider the pricing of the policies. Table~\ref{table:bmk-losses} shows the expected and simulated losses for the distributions in Figure~\ref{fig:bmk-joint-distro}. Note that \[\text{Expected Loss} = \text{Expected Frequency} \times \text{Expected Severity} \times \text{Number of polices}\]
The ratio of the Expected Loss and the Exposure (\$100mn in this example) gives what is known in insurance as the technical premium rate. Accordingly, the technical premium would be 5\% for the 10\% frequency scenario and 25\% for the 50\% frequency scenario. The simulated losses are lower than the expected (mean) losses because of the skew of the log-normal distribution.  
\begin{table}[htb]
    \centering
    \begin{tabular}{c|cc}
         \hline
         Frequency & Expected Loss & Simulated Loss\\
         \hline
         10\% & \$5mn & \$4.6mn \\
         50\% & \$25mn & \$22.9mn \\
         \hline 
    \end{tabular}
    \vspace{1mm}
    \caption{Expected versus simulated benchmark distribution losses}
    \label{table:bmk-losses}
\end{table}

\begin{figure}[htb]
\centering
\includegraphics[width=16cm]{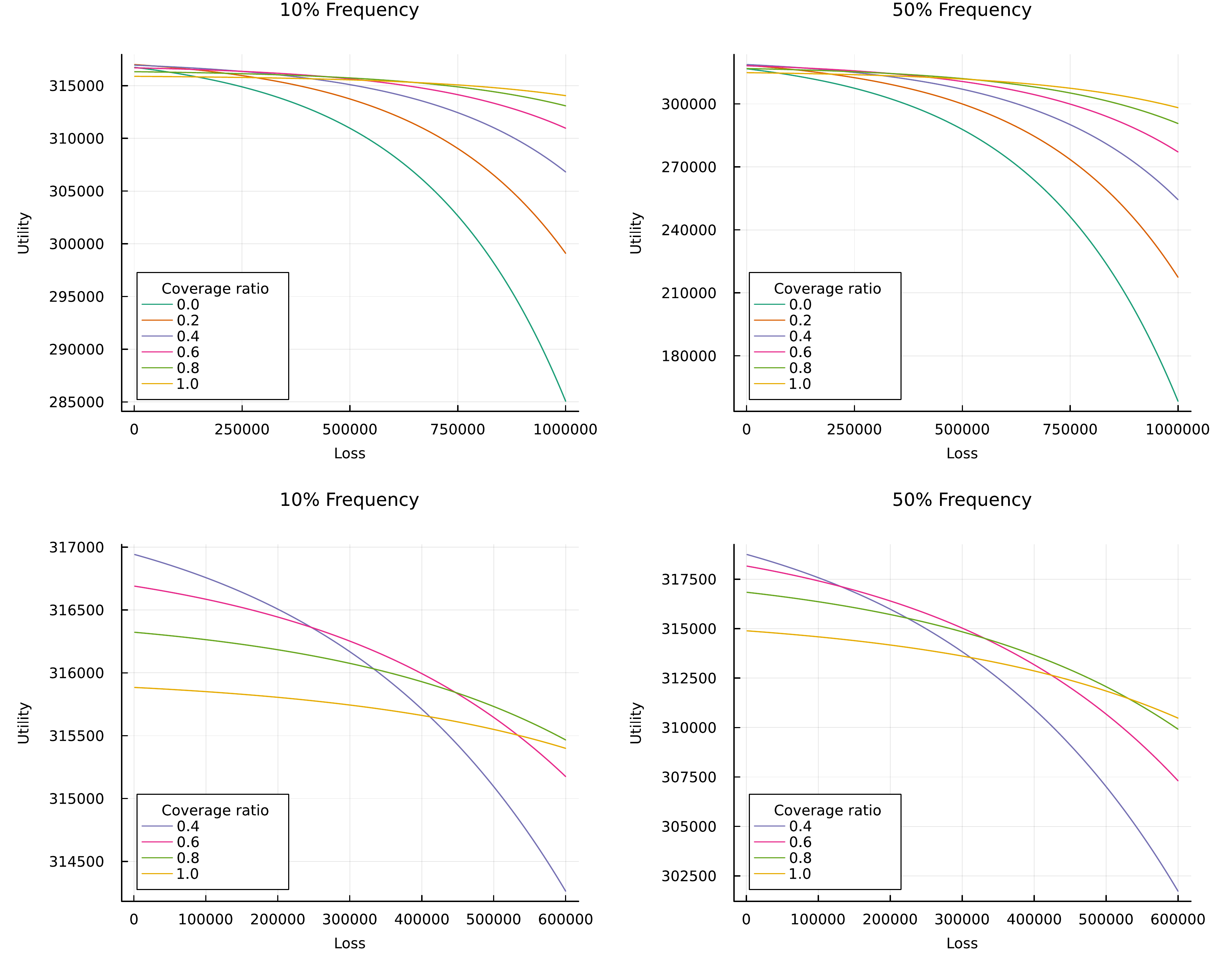}
\caption{Buyer utilities}
\label{fig:bmk-buyer-u}
\end{figure}
Figure~\ref{fig:bmk-buyer-u} plots utilities for insurance buyers (see Equation~\ref{eq:baseline-buyer-utility}) with the constant absolute risk aversion parameter, $\alpha$ fixed at $3 \times 10^{-6}$. This is linked to the loss limit of the single contract for this simulation (\$1mn). $\pi=0.1 \text{ or } 0.5$. The upper graphs in the Figure show the utility for different coverage ratios (these are fractions of the \$1mn limit insured). The lower graphs magnify interesting features of the upper graphs. Based on the premium rates charged, a rational agent would not take full insurance cover at the expected average loss per contract (\$50k for 10\% frequency and \$250k for 50\% frequency) as full coverage does not maximise utility at these loss values. 

To simulate reinsurance pricing, we first fit a log-normal distribution to the joint distribution with 50\% loss frequency as previously described. We consider reinsurance only for the 50\% loss frequency distribution as guided by the reported loss ratios in Table~\ref{table:naic-figures}, which suggest relatively high frequencies of losses have been experienced by the actual market. Using the \textit{fit} functions in \textit{Distributions.jl}, we obtain a log-normal Distribution with $\mu=16.9$ and $\sigma=0.27$. Under this distribution, the cumulative probability of a loss exceeding \$50mn is extremely small, therefore we price the reinsurance policies using excess-of-loss above a varying threshold up to a limit of \$50mn. Using the cumulative probability functions for the estimated distribution, we can then obtain premium rates for the reinsurance, which, multiplied by the amount of reinsurance required, gives the cost of reinsurance. We then re-run the simulations of losses for the insurer assuming no losses are incurred above the threshold at which reinsurance cover binds. We can then obtain the simulated loss with reinsurance and the commensurate simulated reduction in losses relative to the initial benchmark of \$22.9mn. The results are presented in Table~\ref{table:benchmark-reinsurance-premia}. 

\begin{table}[htb]
    \centering
    \begin{tabular}{c|M{2cm}|M{2cm}|M{2.5cm}|M{3cm}|M{2cm}}
         \hline
         Reinsurance&Reinsurance Premium Rate&Reinsurance Cover&Technical Reinsurance Premium&Simulated Loss with Reinsurance&Simulated Loss Reduction  \\
         \hline
         \$25mn xs \$25mn&32.2\%&\$25mn&\$8.1mn&\$13.1mn&\$9.8mn\\
         \$20mn xs \$30mn&12.5\%&\$20mn&\$3.8mn&\$18.7mn&\$4.2mn\\
         \$15mn xs \$35mn&4.2\%&\$15mn&\$0.6mn&\$21.3mn&\$1.6mn\\
         \$10mn xs \$40mn&1.3\%&\$10mn&\$0.1mn&\$22.4mn&\$0.5mn\\
         \hline
    \end{tabular}
    \vspace{1mm}
    \caption{Reinsurance premia for excess-of-loss policies on the 50\% chance of claim distribution}
    \label{table:benchmark-reinsurance-premia}
\end{table}

This reinsurance pricing may be considered efficient because both the reinsurer providing coverage and the insurer seeking reinsurance have the same expected loss distribution. The simulated loss reduction exceeds the cost of reinsurance for each and every of the four reinsurance contracts priced. This likely reflects a small margin of error in fitting the log-normal distribution to represent the joint frequency-severity distribution relative to the underlying distribution of losses which are fully captured by the simulations.   

\subsection{Simulation 2: Reinsurance supply and price}
\label{subsec:re-supply-price}
Having considered the case where all parties agree on the same distribution, we relax this assumption and start to consider divergence in distributions of expected losses. We begin by considering the objective of the reinsurer. We assume a log-normal distribution of total losses. This is the distribution the reinsurance company believes represents the losses experienced from a pool of cedents. The reinsurance company needs to model different potential loss ratios. Initially, we assume cover is fixed at a maximum of \$500mn. Table~\ref{table:reinsurance-distros} presents a number of log-normal distributions. These are purely for illustrative purposes; in a real world situation, the reinsurer would model the distribution based on experience and data. However, it is helpful to consider a range of distributions to understand how the shape of the distribution may affect pricing.     

\begin{figure}[tb]
\begin{center}
\includegraphics[width=16cm]{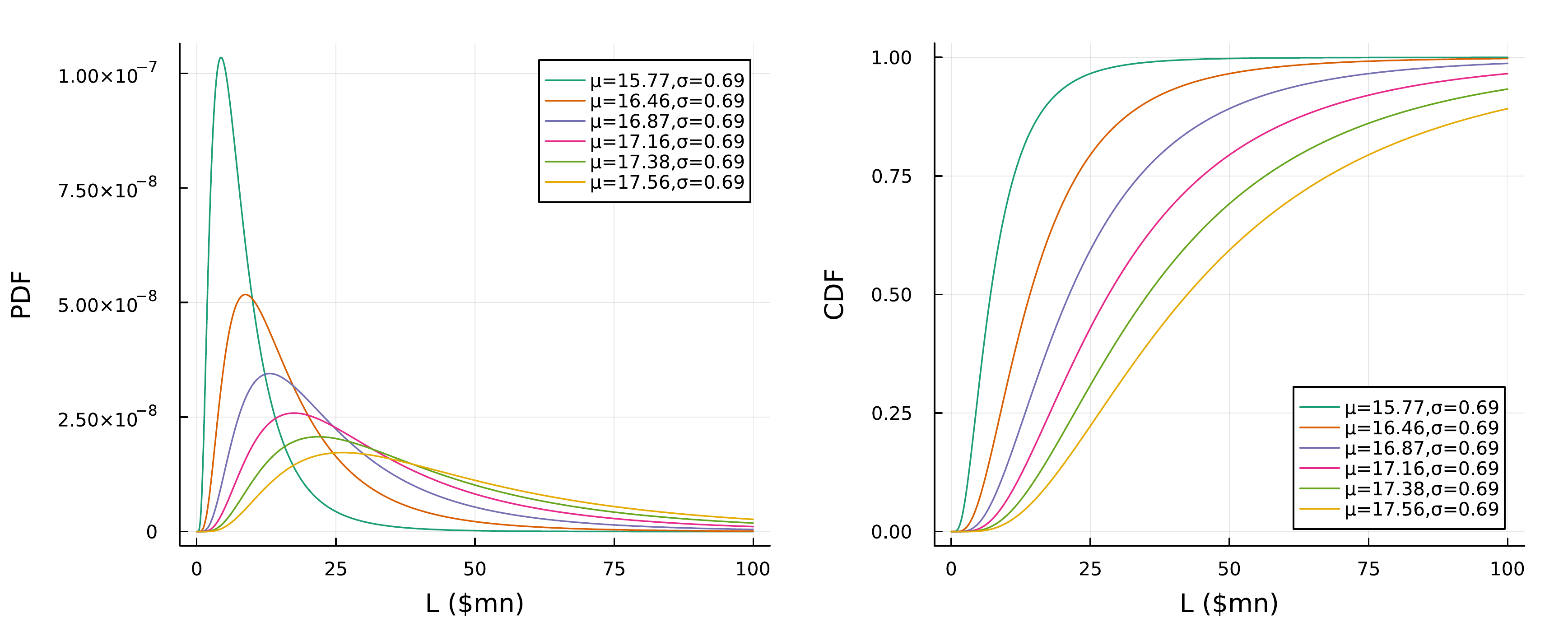}
\end{center}
\caption{\label{fig:reinsurer-ln-distros}Reinsurer loss distributions}
\end{figure}

\begin{table}[H]
\begin{center}
    \begin{tabular}{|l|r|c|c|c|c|c|}
    \hline 
     \multicolumn{2}{|l|}{\rule{0pt}{1em}}&$\mu_L$ & $\sigma_L$ & $\mu$ & $\sigma$ & $F^{-1}(0.995)$ \\
    \hline
   \parbox[t]{2mm}{\multirow{6}{*}{\rotatebox[origin=c]{90}{Distributions}}}&A&\$10mn&\$10mn&15.8&0.69&\$42mn\\
&B&\$20mn&\$20mn&16.5&0.69&\$84mn\\
&C&\$30mn&\$30mn&16.9&0.69&\$126mn\\
&D&\$40mn&\$40mn&17.2&0.69&\$169mn\\
&E&\$50mn&\$50mn&17.4&0.69&\$211mn\\
&F&\$60mn&\$60mn&17.6&0.69&\$253mn\\
    \hline
    \end{tabular}
    \vspace{1mm}
\end{center}
\caption{Table of reinsurance distributions}
\label{table:reinsurance-distros}
\end{table}
Within this table, $F^{-1}(0.995)$ represents the maximum loss with 99.5\% certainty within the distribution. This is the probability value used under the Solvency II insurance regulation to determine the required capital a firm must hold. The probability density function and cumulative distribution functions for the distributions in Table~\ref{table:reinsurance-distros} are plotted in Figure~\ref{fig:reinsurer-ln-distros}. Note that the scale of the loss axis is shortened to \$100mn as the probability density function returns extremely low values beyond this point. 

To estimate the premium rate, we consider the following. The reinsurer targets a loss ratio (a common performance metric in the insurance industry). Total losses from the portfolio are then written
\begin{equation}
    L = \mathrm{L.R.} \times \sum_Jr_JC_J
\end{equation}
Losses experienced are also given by
\begin{equation}
 L = \sum_JE[I_J] 
\end{equation}
We assume there is a single rate for reinsurance such that $r_J =r \forall J$. Then,
\begin{equation}
    r = \frac{\sum_JE[I_J]}{\mathrm{L.R.}\times\sum_J{C_J}}
\end{equation}
Denoting $\bar{C} = \sum_j C_J$ and noting that $\sum_JE[I_J] = \int_0^{\bar{C}} If(I)dI$ where $f(I)$ is the probability density function of an appropriate distribution, we obtain
\begin{equation}
    r = \frac{\int_0^{\bar{C}}If(I)dI}{\mathrm{L.R.}\times\bar{C}}
\end{equation}
This integral can be evaluated numerically, for example using \textit{QuadGK} in Julia. 

\begin{table}[htb]
\begin{center}
    \begin{tabular}{|l|r| c * {8} c|}
    \hline
    \multicolumn{2}{|l|}{}&\multicolumn{9}{|c|}{Premium rates}\\
    \hline
    \multicolumn{2}{|l|}{Loss ratios$\rightarrow$}& 0.1 & 0.2 & 0.3 & 0.4 & 0.5 & 0.6 & 0.7 & 0.8 & 0.9 \\
    \hline

\parbox[t]{2mm}{\multirow{6}{*}{\rotatebox[origin=c]{90}{Distributions}}}&A&0.18&0.09&0.06&0.04&0.04&0.03&0.03&0.02&0.02\\
&B&0.36&0.18&0.12&0.09&0.07&0.06&0.05&0.04&0.04\\
&C&0.54&0.27&0.18&0.13&0.11&0.09&0.08&0.07&0.06\\
&D&0.72&0.36&0.24&0.18&0.14&0.12&0.1&0.09&0.08\\
&E&0.9&0.45&0.3&0.22&0.18&0.15&0.13&0.11&0.1\\
&F&1.08&0.54&0.36&0.27&0.22&0.18&0.15&0.13&0.12\\
\hline
    \end{tabular}
    \vspace{2mm}
\caption{Illustrative premium rates for target loss ratios under different distributions at cover fixed at \$500mn}
\label{table:loss-ratios-fixed-cover}

\end{center}
\end{table}

Suppose the reinsurer believes that Distribution C best describes expected losses to the portfolio and targets a loss ratio of 50\%. The rate of reinsurance charged is then 11\% (Table~\ref{table:loss-ratios-fixed-cover}). Premium income for the reinsurer will be \$55mn. Note that per Table~\ref{table:reinsurance-distros}, in Distribution C, the 99.5\% upper bound for losses is \$126mn. The reinsurer must therefore hold capital of \$71mn under this policy scenario.  

\subsubsection{Insurance supply}

We assume for simplicity that there are five insurance contracts in the market with different limits: \$500k, \$1mn, \$2mn, \$5mn, \$10mn. We assume that there is a uniform individual and independently distributed probability of loss for each contract:
\begin{table}[htb]
\begin{center}
    \begin{tabular}{c|c|c|c|M{2cm}|M{2.5cm}}
         \hline
         Limit& $\mu_L$ & $\sigma_L $ & Frequency ($\pi_L$) & Expected Loss ($\pi_L.\mu_L$) & Premium (Exp. Loss/Limit) \\
         \hline
         \$500k & \$200k & \$125k & 0.1 & \$20k & 4\%\\
         \$1mn & \$400k & \$350k & 0.15 & \$60k & 6\%\\
         \$2mn & \$1mn & \$1mn & 0.16 & \$160k & 8\%\\
         \$5mn & \$2.5mn & \$1.25mn & 0.2 & \$500k & 10\%\\
         \$10mn & \$4mn & \$4mn & 0.3 & \$1.2mn & 12\%\\
         \hline
    \end{tabular}
\vspace{2mm}
\caption{Insurance contracts in the market}
\label{table:market-insurance-contracts}
\end{center}
\end{table}
The expected severity in the above contracts is assumed to be log-normally distributed per Table~\ref{table:market-insurance-contracts} and the frequency $\sim\mathrm{Poisson}(\pi_l k)$ where $k$ is the number of contracts. Table~\ref{table:policies-written} contains a sample portfolio for a panel of 5 insurers for illustrative purposes to run a loss simulation. The technical premium is the premium income that equates to the expected loss. Equivalently, this is the premium written at which the insurer would expect to break even. 
\begin{table}[htb]
\begin{center}
    \begin{tabular}{c| *{5}{c}|c|c}
    \hline
    & \multicolumn{5}{c|}{Policy count grouped by policy limit} && \\
    \hline
    Insurer & \$500k & \$1mn & \$2mn & \$5mn & \$10mn & Total Exposure & Technical Premium \\
    \hline
    Alpha & 200 & 0 & 0 & 0 & 0 & \$100mn & \$4.0mn \\
    Beta & 100 & 50 & 0 & 0 & 0 & \$100mn& \$5.0mn\\
    Charlie & 50 & 20 & 15 & 5 & 0 & \$100mn & \$7.1mn\\
    Delta & 30 & 0 & 5 & 5 & 5 & \$100mn & \$9.9mn\\
    Echo & 0 & 0 & 0 & 0 &  10 & \$100mn& \$12.0mn\\
    \hline
    Total & 380 & 70 & 20 & 10 & 15 & \$500mn & \$38.0mn \\
    \hline
    \end{tabular}
\end{center}
\caption{Insurance policies written by insurance panel}
\label{table:policies-written}
\end{table}

In reality, insurers do not attempt to break even but rather aim to produce a profit to provide a return on investment to the source of their capital. One simple objective might to not exceed a target loss ratio. This is achieved via an additional charge to the insurance buyer over the actuarial fair premium known as a loading\footnote{See, for example, Benjamin (1986)~\cite{benjamin1986loadings} for a discussion.}. The loading is calculated:
\begin{equation}
    \begin{aligned}
   & \mathrm{Loading} = \\ &\frac{1}{\mathrm{Total\ Exposure}} \times\biggl( \frac{\mathrm{Technical\ Premium}}{\mathrm{Target\ Loss\ Ratio}}-\mathrm{Technical\ Premium}\biggl)
    \end{aligned}
\end{equation}

\begin{figure}[!htb]
\centering
\includegraphics[width=12cm]{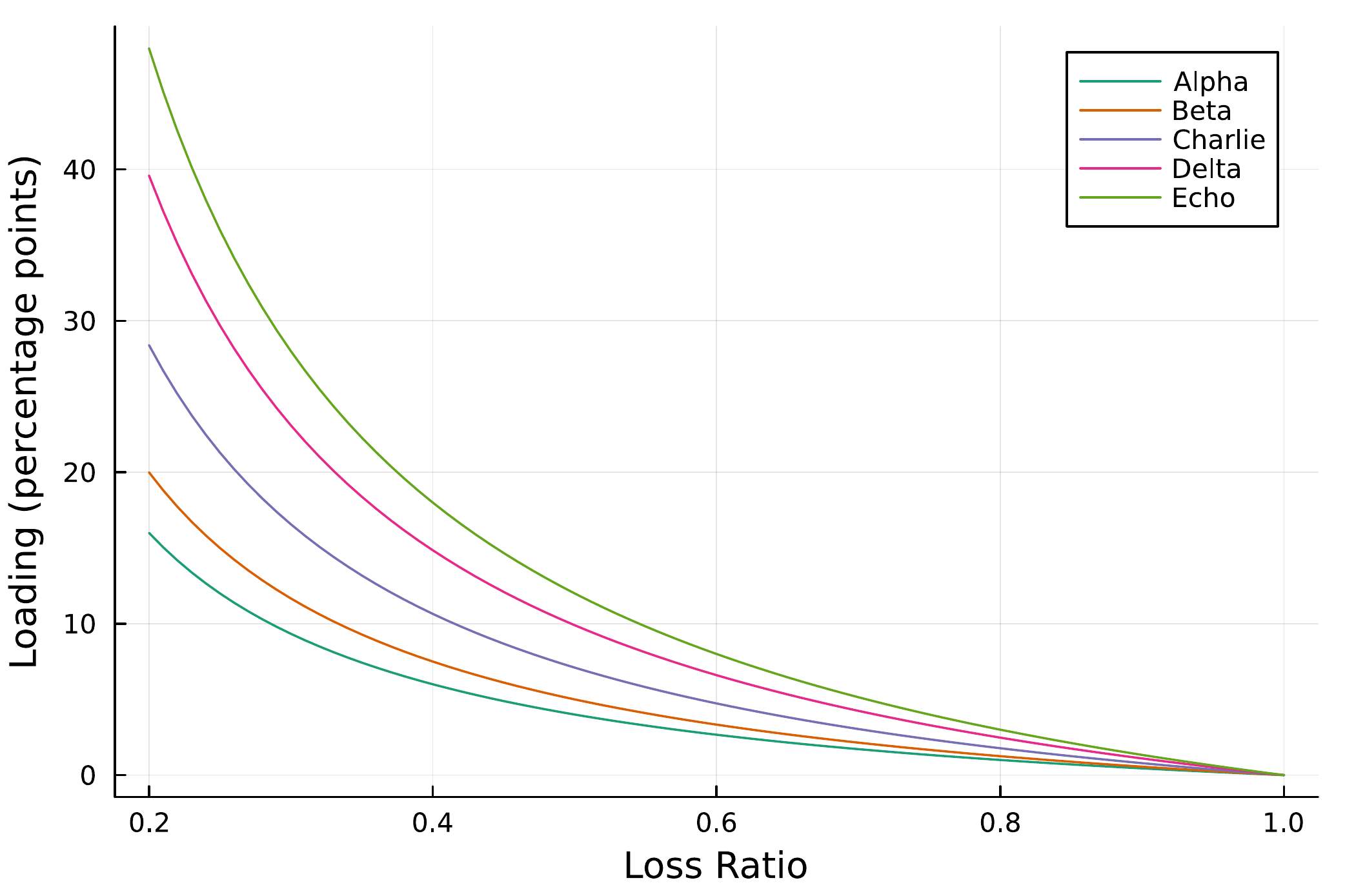}
\caption{Loading versus target loss ratio for different insurance portfolios}
\label{fig:insurer-profitability-curve}
\end{figure}

The variation between loading and loss ratio for the insurance portfolios in Table~\ref{table:policies-written} is plotted in Figure~\ref{fig:insurer-profitability-curve}. The variation in target loss ratios may occur for a number of reasons, such as rate of return on capital demanded by the capital source (as discussed in Section~\ref{subsec:market-structure}, prior loss experience, or other variable expenses. The loading also may aim to capture any skew in the actuarial distribution.

Table~\ref{table:loading-factors} shows the calculated loadings for each insurer in the simulation assuming a target loss ratio of 50\%. For ease of comparison, we keep the target loss ratio constant across the insurer panel and also the overall exposure. 

\begin{table}[htb]
    \begin{center}
        \begin{tabular}{M{1.5cm}|M{2cm}|M{2cm}|M{2cm}|M{2cm}|c|M{2cm}}
             \hline
             Insurer & Technical Premium & Target Loss Ratio & Exposure&Technical Premium Rate &Loading&Weighted Average Charged Premium Rate \\            
             \hline
             Alpha  & \$4.0mn & \hfil 50\% & \$100mn & 4.0\% & 4.0pp&\hfil 8.0\%\\
             Beta  & \$5.0mn & \hfil 50\% & \$100mn & 5.0\%&5.0pp& \hfil 10.0\%\\
             Charlie  & \$7.1mn & \hfil 50\% & \$100mn & 7.1\%& 7.1pp& \hfil 14.2\%\\
             Delta  & \$9.9mn & \hfil 50\% & \$100mn & 9.9\%& 9.9pp& \hfil 19.8\%\\
             Echo  & \$12.0mn & \hfil 50\% & \$100mn & 12.0\%& 12.0pp& \hfil 24.0\%\\
             \hline
              
        \end{tabular}
    \end{center}
\caption{Calculating premium loading rates for insurance companies based on simulated losses. The loading rate is expressed in percentage points.}
\label{table:loading-factors}
\end{table}

\subsubsection{Interaction between insurance and reinsurance}
The total expected losses for the cyber-insurance market depicted in Table~\ref{table:loading-factors}  are \$38.0. The technical premium is equal to the expected losses in monetary terms. In Section~\ref{subsec:re-supply-price} we stated that for a log-normal distribution with mean and standard deviation of \$40mn and target loss ratio of 0.5, the premium charge would be 14\% for the reinsurer (distribution D). The usual process of reinsurance in quota share is that the reinsurer assumes a stated percentage of portfolio losses. The reinsurance contract (or treaty) is priced\footnote{Clark (2014)~\cite{clark2014} is a highly approachable introducing to reinsurance pricing} via a ceding commission and reinsurance margin. In this case, the reinsurance margin is already accounted for in the 14\% premium rate as this was calculated to give the required reinsurer loss ratio. The ceding commission is paid back to the ceding insurer to compensate them for underwriting expenses. The ceding commission is defined as the average premium rate (Table~\ref{table:loading-factors}) less the cost of reinsurance (14\%). Inspecting Table~\ref{table:loading-factors} once more, we can see for insurers Charlie, Delta, and Echo, the average premium rate of the portfolio exceeds the reinsurance cost. Therefore, the ceding commission for these insurers would be positive. However, for insurers Alpha and Beta, their weighted average premium rate is below that charge for reinsurance, implying a negative ceding commission. If Alpha or Beta believe that their assumed distributions are correct, this would not be rational behaviour. For the other insurers, purchasing reinsurance would reduce profits for \emph{expected losses}. However, the value of reinsurance will become apparent once we consider the effect of capital. 

Having established the target pricing for each insurer \textit{ex ante}, we now consider simulating \textit{ex post} losses. The profit equation for the insurer, may be written:
\begin{equation}
    \begin{aligned}
    \mathrm{\$Profit}(L) &= \mathrm{\$Premium\ Written}\times(1-\rho) \\&+(\mathrm{\$Exposure}\times\rho\times\mathrm{\%Ceding\ Commission)} \\ &-L
    \end{aligned}
\label{eq:profit-with-re}
\end{equation}
\begin{equation}
    L =
    \begin{cases}
    \text{Loss}(1-\rho) & \text{if } D =0\\
    \text{Loss} & \text{if } D > 0, \text{Loss} \leq D\\
    D + (1-\rho)(\text{Loss}-D) & \text{if } D > 0, \text{Loss} > D
    \end{cases}
\end{equation}
$\rho$ is the fraction of the portfolio ceded to the reinsurer and $D$ is a deductible. We restrict our analysis in this simulation solely to policies without deductibles, but provide for their inclusion for completeness. 

\subsubsection{Simulation Procedure} \label{subsubsec:simul-proc}
For each insurance portfolio in Table~\ref{table:market-insurance-contracts} we simulate losses via the following procedure.
\begin{enumerate}
    \item Set severity distribution for each contract as in Table~\ref{table:market-insurance-contracts}.
    \item Set frequency distribution as per Table~\ref{table:market-insurance-contracts} --- $\text{Poisson} \sim \pi_L . k$ where $k$ is the number of each contract contained in the portfolio.
    \item Randomly sample the frequency of expected losses for each contract in the portfolio, to generate a number of losses for \emph{each contract}, $N_{\text{loss}}$.
    \item Randomly sample from the severity distribution for each contract $N_{\text{loss}}$ times, sum and record the losses.
    \item Run the above process 100,000 times.
\end{enumerate}
The results of the simulations are presented in Table~\ref{table:simulated-losses-ins-ports} (histograms of the generated loss distributions are provided in Appendix~\ref{appendix}). The table contains the premium income for each insurer as previously determined, a capital level assumed to be held by the insurer equal to the average baseline loss in the simulation and reserves defined, 
\begin{equation}
    \text{Reserves} = \text{Premium Written}+\text{Capital}
\end{equation}
Along with the simulated loss values, we also calculate loss values for a `stress test' type scenario, calculating the maximum loss in 95\% and 97.5\% of cases. This is done via using the \textit{quantile} function of \textit{Distributions.jl} to calculate the respective frequency and severity at $F^{-1}(0.95)$ and $F^{-1}(0.975)$. The required values are then readily obtained. With these values obtained, we may now proceed to consider the interaction between reinsurance and the insurance portfolios. 

\begin{table}[htb]
    \centering
    \begin{tabular}{c|M{2cm}|c|c|M{2cm}|M{2cm}|M{1.5cm}|M{1.5cm}}
    \hline
        &\multicolumn{3}{c|}{Assets}&\multicolumn{4}{c}{Losses}\\
        \hline
         Insurer&Premium Income&Capital&Reserves&Simulation Baseline Average&Simulation Baseline SD&95\% Stress Test&97.5\% Stress Test\\
         \hline
Alpha&\$8.0mn&\$3.6mn&\$11.6mn&\$3.6mn&\$0.8mn&\$8.2mn&\$9.4mn \\
Beta&\$10.0mn&\$4.4mn&\$14.4mn&\$4.4mn&\$1.3mn&\$13.6mn&\$17.4mn \\
Charlie&\$14.2mn&\$6.4mn&\$20.6mn&\$6.4mn&\$3.0mn&\$28.0mn&\$36.6mn \\
Delta&\$19.8mn&\$8.9mn&\$28.7mn&\$8.9mn&\$6.2mn&\$51.2mn&\$64.9mn \\
Echo&\$24.0mn&\$10.8mn&\$34.8mn&\$10.8mn&\$7.9mn&\$53.1mn&\$77.0mn \\
        \hline
        
    \end{tabular}
    \vspace{1mm}
    \caption{Simulated Losses}
    \label{table:simulated-losses-ins-ports}
\end{table}

\subsubsection{Considering the effect of capital}
Suppose, as per Table~\ref{table:simulated-losses-ins-ports} that the insurer has a capital buffer, which initially, is equal to the simulated average losses on its portfolio. We now examine the optimal reinsurance fraction which means the insurer would remain solvent in the event of losses of a specified magnitude. We consider $\rho$ values for both the 95\% and 97.5\% stress tests. This means calculating the value of $\rho$ which would set $\text{\$Profit}(L) = -K$ (Equation~\ref{eq:profit-with-re}). The required expression is

\begin{equation}
\begin{aligned}    
    &\bar{\rho}(L_{\text{stress}}) = \\ & \frac{(L_{\text{stress}} -\text{\$Premium Written}-K)}{L_{\text{stress}}-\text{\$Premium Written}+(\text{\$Exposure}\times\text{\%CC})}
\end{aligned}
\end{equation}
where $\%CC$ is the percentage ceding commission. 

\begin{figure}[htb]
\centering
\includegraphics[width=16cm]{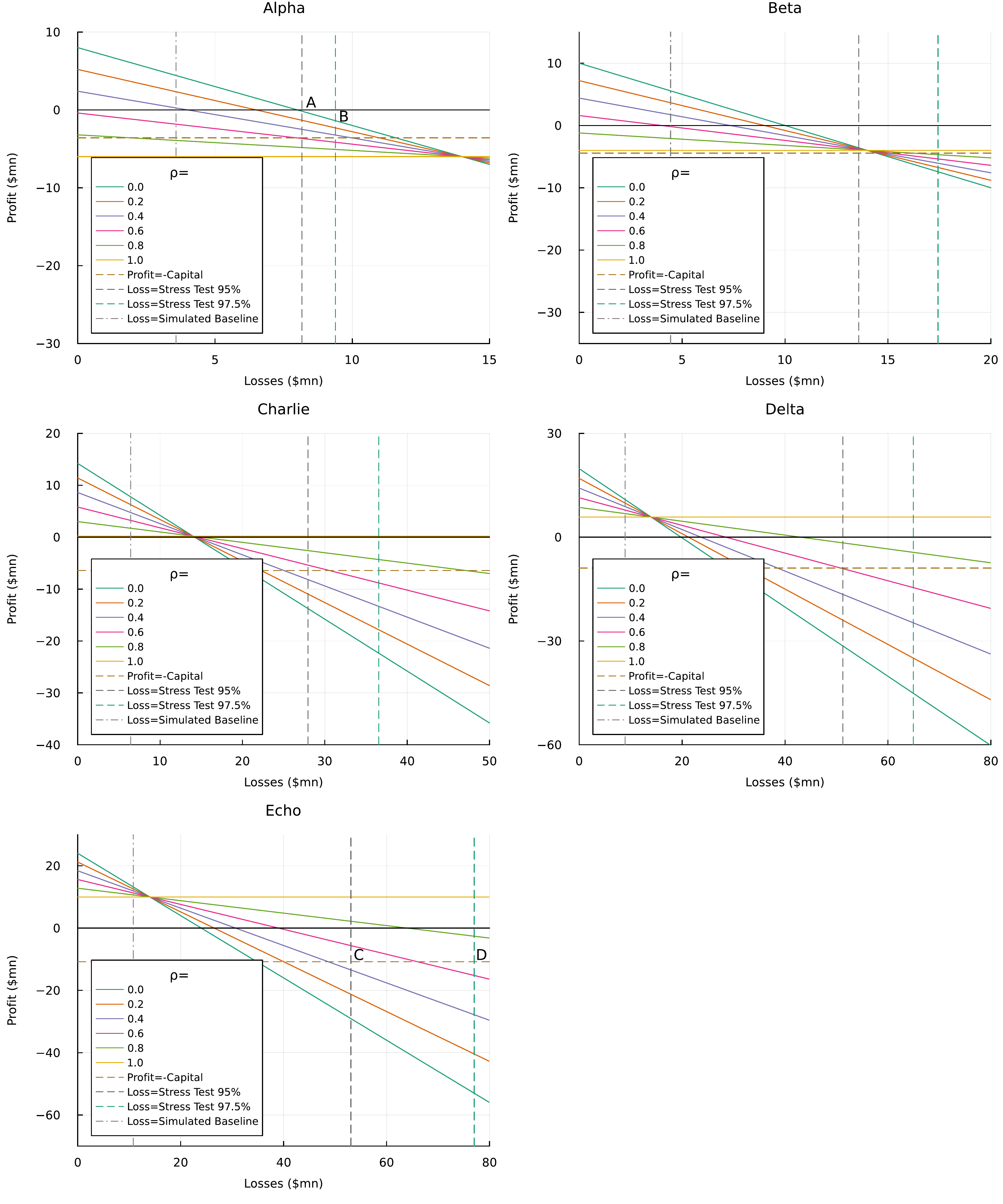}
\caption{Insurer profit with varying quota share proportions ($\rho$)}
\label{fig:ins-re-pl-no-ded}
\end{figure}

The solvency threshold for the insurer is $\text{Reserves}=L_{stress}$. If $\text{Reserves}>L_{stress}$ then we set $\bar{\rho}=0$ as the insurer does not need reinsurance at this stress test loss level as it would remain solvent without it. For insurer Alpha, reserves exceed the stress test losses at both thresholds, while for Beta, reserves exceed only the 95\% stress test loss. Figure~\ref{fig:ins-re-pl-no-ded} and Table~\ref{tab:rho-optimal} show the complete results of the analysis. Starting with Table~\ref{tab:rho-optimal}, it appears that neither Alpha nor Beta should buy reinsurance. In the 97.5\% stress-test, Beta is insolvent even with reinsurance. This suggests that Beta would need to implement a higher loading than that initially calculated to pass the stress test. For Charlie, Delta, and Echo, there is benefit in purchasing reinsurance as a quota share policy, though the optimal fractions appear fairly high. Consequently, the insurers might decide to buy less than the optimum but set capital higher. However, this then means that the market is not efficient. Table~\ref{tab:rho-optimal} also shows the profit each insurer would receive if \textit{ex post} losses equal the simulated baseline with no reinsurance; with reinsurance at the $\rho^{0.95}$ fraction; and with reinsurance at the $\rho^{0.975}$ fraction. For Charlie, given that its weighted average premium rate is close to the reinsurer objective, it receives a scant ceding commission. Consequently, there is an opportunity cost of \$4.0-5.2mn of purchasing quota share at the optimum relative to baseline simulated profit of \$7.8mn. In a market where information is shared, there should not be an opportunity cost. For Delta and Echo, the purchase of quota share appears more attractive because of the more generous ceding commission. These are deliberately extreme examples, but in practice suggest that bargaining may occur between different insurers and reinsurers over the ceding commission, which introduces inefficiency into the market. 

\begin{table}[htb]
    \centering
    \begin{tabular}{c|M{3cm}|c|c|M{2cm}|M{2cm}|M{2cm}}
        \hline
        \multicolumn{4}{c|}{}& \multicolumn{3}{c}{Profit if Losses=Simulation Baseline (\$mn)} \\
         \hline
         Insurer & Ceding Commission&$\bar{\rho}^{0.95}$ & $\bar{\rho}^{0.975}$&$\rho=0$& $\rho=\bar{\rho}^{0.95}$& $\rho=\bar{\rho}^{0.975}$\\
         \hline
Alpha&-6.0\%&0.00&0.00&4.4&4.4&4.4 \\ 
Beta&-4.0\%&0.00&0.87&5.6&5.6&-2.7 \\ 
Charlie&0.2\%&0.53&0.71&7.8&3.8&2.4 \\ 
Delta&5.8\%&0.60&0.71&10.9&7.8&7.3 \\ 
Echo&10.0\%&0.47&0.67&13.3&11.7&11.1 \\ 
         \hline 
    \end{tabular}
    \vspace{1mm}
    \caption{Reinsurance ceding fractions that maintain insurer solvency at different stress-test values}
    \label{tab:rho-optimal}
\end{table}
Figure~\ref{fig:ins-re-pl-no-ded} presents a more detailed picture of the simulations that yield the optimal $\rho$. For each insurer portfolio, we plot the insurer profit (Equation~\ref{eq:profit-with-re}) as a function of losses for values of $\rho$ between 0 and 1. The capital held (i.e. the average simulated loss as already discussed) is represented as a horizontal line and the average simulated losses for the 95\% and 97.5\% stress tests are represented as vertical lines. The intersection of the average simulated loss and the stress test allows for the optimal \$rho to be read from the graphs. In the case of Alpha, it can be seen that on the $\rho=0$ profit line, at the two stress test loss values (Points A and B), the profit exceeds the capital held. For reinsurance to be worth purchasing, the $\rho=0$ profit line must be less than the capital horizontal lines at the stress test losses. Taking Echo as an example, with stress test loss of 95\%, we can see that the horizontal capital and vertical loss lines intersect between the profit lines for $\rho=0.4$ and $\rho=0.6$ (Point C) As may be verified from Table~\ref{tab:rho-optimal}, the reinsurance fraction for this case is 0.47. The comparable intersection for the 97.5\% stress test (Point D) is at $\rho=0.67$. 

\subsubsection{Excess of Loss}
Having considered the quota share case, it is worth considering the case of excess-of-loss insurance as an alternative to quota share for the insurers. Rather than using the capital buffer approach, we consider a simpler objective: that the insurer rather than holding a capital buffer buys insurance from a reinsurer to cover losses in excess of its cash premium income up to the limit of the 97.5\% stress test loss value. To calculate the required parameters, we can use the simulated baseline losses already calculated in Section~\ref{subsubsec:simul-proc}. From these, we compute the number of instances of losses in the vector of generated losses that exceed the cash premium income but are less than the 97.5\% stress test loss value. 

\begin{table}[bt]
    \centering
    \begin{tabular}{c|M{3.5cm}|M{2.5cm}|M{2.5cm}|M{2.5cm}}
         \hline
         Insurer&XL Reinsurance Coverage&Probability of Loss $>$ Cash Premium&Technical XL Reinsurance Premium&QS Reinsurance Premium at $\bar{\rho}^{0.975}$\\
         \hline 
Alpha & \$1.4mn xs \$8.0mn & 0.0\%&\$0.0mn&\$0.0mn\\ 
Beta & \$7.4mn xs \$10.0mn & 0.0\%&\$0.0mn&\$8.3mn\\ 
Charlie & \$22.4mn xs \$14.2mn & 1.6\%&\$0.4mn&\$5.4mn\\ 
Delta & \$45.1mn xs \$19.8mn & 5.9\%&\$2.7mn&\$3.6mn\\ 
Echo & \$53.0mn xs \$24.0mn & 6.6\%&\$3.5mn&\$2.2mn\\ 
\hline
    \end{tabular}
    \vspace{1mm}
    \caption{Excess of Loss Pricing Example}
    \label{table:excess-of-loss-ins-re}
\end{table}

The results are contained in Table~\ref{table:excess-of-loss-ins-re}. The portfolios of Alpha and Beta generate expected losses well below the level of cash premium income (see Appendix~\ref{appendix}) and accordingly there is little benefit in excess-of-loss insurance. For Charlie, Delta, and Echo, it is interesting to note that the combined technical premium is \$7mn. Recall that in Section~\ref{subsec:re-supply-price}, Distribution A in Table~\ref{table:loss-ratios-fixed-cover} gives the loss ratios for the reinsurer versus quoted premium for an expected \$10mn of losses. If we assume that the reinsurer requires a loss ratio of 0.5 or better, then the minimum premium it will charge is 4\%. For the insurance buyers, only for Echo is buying excess-of-loss reinsurance cheaper than buying quota share. Consequently, for each insurance portfolio, there is a different optimal reinsurance contract from the perspective of the insurance company seeking reinsurance.

The excess-of-loss premia calculated in Table~\ref{table:excess-of-loss-ins-re} are computed using the individual joint distributions of frequency and severity for each of the five insurance companies. These are known only to each of those insurance companies alone and are not visible to the reinsurer. Consequently, there are information asymmetries between the insurers seeking reinsurance and the reinsurer. Insurers Delta and Echo know that the fair insurance premium rate for the excess-of-loss contracts specified in Table~\ref{table:excess-of-loss-ins-re} are 5.9\% and 6.6\%, respectively. However, the reinsurer would offer these contracts at 4\% premium rate based on its own distribution. Consequently, the insurers can, under these assumptions, buy reinsurance cheaper than its fair cost based on their avantageous knowledge of the `true' distribution rather than the reinsurer's distribution which assumes simple log-normal distribution of a set of risks at a particular expected loss value. This illustrates how inefficiency and therefore financial imbalances between insurance and reinsurance may emerge as a consequence of different expected loss distributions, unlike in Table~\ref{table:benchmark-reinsurance-premia} where the reinsurer and insurer(s) had the same distribution of expected losses. 

\subsection{Simulation 3: Insurance buyers of variable risk}

We now consider a simulation in which buyers have heterogeneous preferences and risk tolerance. The interactions of the real insurance market are hard to model as insurance customers interact with insurance companies via insurance brokers who act as an intermediary. The flow of business is directed therefore partly by relationships (and so is not efficient in a traditional economic sense). However, it is possible to construct some simulations of insurance demand based on different characteristics and illustrate the utility demand model and how this may affect reinsurance pricing. 

The insurance buyer faces a single utility maximization decision: for a given premium rate, how much cover does the agent with to purchase. This could be formalised in terms of expected utility (Equation~\ref{eq:baseline-buyer-utility-state-dep}) via variation of the risk aversion parameter, $\alpha$, but this is not necessary for the example presented here. The insurance company must choose premium rates that it believes will not excessively deplete its capital for a certain level of risks, or plan to cede premium to reinsurance to cover that risk as demonstrated in the previous section. We will retain the contract limit structure from Table~\ref{table:market-insurance-contracts} for this analysis, meaning that insurance buyers choose one of the five contracts. 

We will now assume that the more coverage the buyer takes, the more sophisticated its assessment of the risks are. This places a constraint on the amount of loading the insurer can apply to the higher limit contracts. We will, as previously, fix the total \textit{potential} cover available in the market at \$500mn and consider how this may be allocated among buyers. However, as will be illustrated, the risks associated with some contracts make them commercially unviable even if theoretically priceable. Table~\ref{table:buyer_restrictions} sets out some arbitrary premia based on the subjective beliefs of the respective buyers, and the maximum number of contracts available in the market based on the overall capacity of \$500mn. We wish to stress that these numbers are established purely for model convenience and to illustrate the further difficulties to establishing efficiency under heterogeneous buyer beliefs. The assumption of market size is required to price potential reinsurance on insurer policies.  

For this analysis, we set the expected severity loss mean equal to a quarter of the policy limit and the standard deviation to half the mean. Unlike in the previous section, we will allow the distribution of expected losses to vary with different clients and have a mixture of buyers considered low-, medium-, and high-risk with different distributions accordingly. We assume that the variation in risk characteristics of the three buyer groups is expressed through variation in frequency.  

\begin{table}[htb]
    \centering
    \begin{tabular}{c|c|c|c|c|c|c}
        \hline
         & \multicolumn{3}{c|}{Highest premium rate at which a buyer takes full coverage}& \multicolumn{3}{c}{Maximum number of customers} \\
         \hline
         Limit & Low risk & Medium risk & High risk & Low risk & Medium risk & High risk \\
         \hline
           \$500k & 14\% & 20\% & 26\% &46&46&46 \\
         \$1mn & 13\% & 18\%&23\%&32&32&32\\ 
         \$2mn & 12\% & 16\%&20\%&16&16&16\\
         \$5mn & 11\% & 14\%&17\%&8&8&8\\
         \$10mn & 10\% & 12\% & 14\%&4&4&4 \\
         \hline
          
    \end{tabular}
    \vspace{1mm}
    \caption{Insurance buyer premium ceilings}
    \label{table:buyer_restrictions}
\end{table}

We assume that reinsurers consider the risks involved for the three different risk categories and apply distributions A, C, and E (Table~\ref{table:reinsurance-distros}) to low, medium and high risks effectively, and target loss ratios of 0.3, 0.5, and 0.7 respectively. This means that the reinsurance charges for the portfolios are 6\%, 11\% and 13\%.

We now consider the distributions associated with the different contracts. Table~\ref{table:contract_spec_buyers} shows the severity and frequency distributions for each policy. We have fixed the severity on each contract and assumed that riskier clients have a higher expected frequency of claims. This assumption could, of course, be varied further, but this approach suffices for the purposes of this example. From this, we simulate the losses with 100,000 runs and derive the expected loss for the entire set of possible contracts. This is shown in Table~\ref{table:buyer_expected_losses} along with the expected average loss per contract. 

\begin{table}[htb]
    \centering
    \begin{tabular}{c|c|c|c|c|c|c}
        \hline
         & \multicolumn{3}{c|}{Severity} & \multicolumn{3}{c}{Frequency, Poisson($\lambda$)}  \\
         \hline
         Limit&$\mu_L$&$\sigma_L$&Distribution&Low risk&Medium risk&High risk\\
         \hline
         \$500k& \$125k&\$62.5k&LogNormal(11.6,0.22)&4.6&11.5&23\\
         \$1mn&\$250k&\$125k&LogNormal(12.3,0.22)&6.4&12.8&19.2\\
         \$2mn&\$500k&\$250k&LogNormal(13.0,0.22)&4&8&12\\
         \$5mn&\$1.25mn&\$625k&LogNormal(13.9,0.22&2&4&6\\
         \$10mn&\$2.5mn&\$1.25mn&LogNormal(14.6,0.22)&1&2&3\\
         \hline
    \end{tabular}
    \vspace{1mm}
    \caption{Distribution specification for insurance contracts offered to buyers}
    \label{table:contract_spec_buyers}
\end{table}

With this calculated, we can then derive the technical premium for each contract, which is shown in Table~\ref{table:fair-premium-buyer}. Comparing with Table~\ref{table:buyer_restrictions}, we can see that for the \$5mn and \$10mn limits, the high risk technical premium is higher than what customers are willing to pay. It may be possible in this case for the insurer to instigate a deductible and reduce the premium. Otherwise, margin is very limited for medium-risk \$5mn and \$10mn limits, which might also motivate introducing a deductible. 

\begin{table}[t]
    \centering
    \begin{tabular}{c|c|c|c|c|c|c}
    \hline
    & \multicolumn{3}{c|}{Expected Loss (Total, \$mn)}&\multicolumn{3}{c}{Expected Loss per Contract (\$k)}\\
    \hline
    Limit & Low risk & Medium risk & High risk & Low risk & Medium risk & High risk\\
    \hline
\$500k& 0.53 & 1.32 & 2.63 & 11 & 29 & 57 \\
\$1mn& 1.47 & 2.93 & 4.40 & 46 & 92 & 138 \\
\$2mn& 1.84 & 3.67 & 5.50 & 115 & 229 & 344 \\
\$5mn& 2.28 & 4.59 & 6.89 & 285 & 574 & 861 \\
\$10mn& 2.30 & 4.59 & 6.86 & 574 & 1,147 & 1,716 \\
    \hline
    \end{tabular}
    \vspace{1mm}
    \caption{Expected losses for policies}
    \label{table:buyer_expected_losses}
\end{table}

\begin{table}[!htb]
    \centering
    \begin{tabular}{c|c|c|c}
    \hline
    & \multicolumn{3}{c}{Technical Premium (\%)}\\
    \hline
    Limit & Low risk & Medium risk & High risk\\
    \hline
  \$500k  & 2.3 & 5.7 & 11.5 \\ 
\$1mn& 4.6 & 9.2 & 13.8 \\ 
\$2mn& 5.7 & 11.5 & 17.2 \\ 
\$5mn& 5.7 & 11.5 & 17.2 \\ 
\$10mn & 5.7 & 11.5 & 17.2 \\ 
\hline
    \end{tabular}
    \vspace{1mm}
\caption{Fair premium for insurance contracts}
\label{table:fair-premium-buyer}
\end{table}

We now consider the capital requirements associated with the insurance policies. Table~\ref{table:buyer_capital_requirements} shows the expected losses for $F^{-1}(0.995)$ and $F^{-1}(0.5)$ for frequency and severity respectively for both the whole set of contracts and also per contract. Each insurer must decide how to allocate its available capital and how much reinsurance to purchase. Rather than calculating sample portfolios, we will simply calculate the reinsurance fraction that is optimal based on Equation~\ref{eq:profit-with-re}.  

\begin{table}[htb]
    \centering
    \begin{tabular}{c|c|c|c|c|c|c}
    \hline
    & \multicolumn{3}{c|}{Stress Test Loss (Total, \$mn)}&\multicolumn{3}{c}{Stress Test Loss per Contract (\$k)}\\
    \hline
    Limit & Low risk & Medium risk & High risk & Low risk & Medium risk & High risk\\
    \hline
\$500k& 1.2&2.3&4.0&27&51&87 \\
\$1mn& 3.1 & 5.1 & 6.9 & 98 & 161 &217 \\
\$2mn& 4.5 & 7.2 & 9.8 & 280 & 447 & 615 \\
\$5mn& 6.7 & 11.2 & 14.5 & 839 & 1,398 & 1,817 \\
\$10mn& 8.9 & 13.4 & 17.9 & 2,236 & 3,354 & 4,472 \\
    \hline
    \end{tabular}
    \vspace{1mm}
    \caption{Stress Test losses for policies, with Frequency set at $F^{-1}(0.995)$, Severity at $F^{-1}(0.5)$}
    \label{table:buyer_capital_requirements}
\end{table}

Based on the Stress Test loss values, and assuming that the insurer writing each contract holds capital equal to the expected value of losses for the contract (Table~\ref{table:buyer_expected_losses}), we can then derive the optimal reinsurance fraction for each contract. As in the prior section (Table~\ref{tab:rho-optimal}), this is calculated by calculating the reinsurance fraction that sets the profit to the insurer equal to $-K$, i.e. at the level of loss given in the Stress Test, the insurer breaks even if it holds this proportion of reinsurance. As the buyers of the smaller contracts are less knowledgeable and will accept a higher premium, the reinsurance fraction is lower as the insurer writes more premium. However, the reinsurance fraction increases from an average of 20\% for the \$500k limit contract to as high as 64\% for the medium-risk \$10mn limit contract. It is clear from this analysis that while it is possible to achieve risk transfer between insurance buyer, insurance company and reinsurer, for a simulated market, achieving convergence of distributions is extremely unlikely as each party is incentivized to maximize their profit rather than target efficiency.  

\begin{table}[htb]
\centering
\begin{tabular}{c|c|c|c}
    \hline
    & \multicolumn{3}{c}{Optimal reinsurance fraction for each contract} \\
    \hline
    Limit & Low risk & Medium risk & High risk \\
    \hline
    \$500k & 0.23 & 0.23 & 0.20 \\
    \$1mn & 0.31 & 0.30 & 0.25 \\
    \$2mn & 0.41 & 0.40 & 0.36 \\
    \$5mn & 0.51 & 0.53 & 0.48 \\
    \$10mn & 0.63 & 0.64 & 0.60 \\
    \hline 
\end{tabular}
\vspace{1mm}
\caption{Optimal reinsurance purchase fraction for each contract implied by stress test values}
\label{table:optimal-re-frac-buyer}
\end{table}

We have stopped short of simulating the allocation of policies to individual insurers as to model competitive market dynamics under uncertainty with heterogeneous beliefs is a complex problem that in itself might fill multiple papers. However, it is hoped that the simulation presented illustrates the additional dynamics that heterogeneous buyer beliefs brings to the challenges of modelling cyber-insurance and re-insurance. To place the simulation results in context with the US cyber-insurance market, in 2020, according to the NAIC~\cite{naic2021}, there were approximately 4 million cyber-insurance policies written in the US market, with the top 20 insurers taking 68\% market share. The report for 2021 does not provide a policy number, but notes that almost 50\% of cyber-insurance premia were ceded.

\section{Discussion} \label{sec:discussion}

The simulations show the difficulty of achieving economic efficiency in an artificial cyber-insurance market even using relatively standard distributions and contract structures. However, as has been stressed, just because a market is not efficient does not mean that transactions cannot take place. We now consider some of the further informational barriers to facilitating smooth transfer of cyber-risk. Issues of data transparency, attack measurement, and reporting --- making relevant data publicly available --- are particularly crucial in enabling agents to make informed pricing decisions. 

\subsection{Information asymmetry}
By and large insurance and reinsurance companies operate in environments where high quality precision signals about loss risks exist. For example, in the case of natural catastrophes, their frequencies are well known and established over many periods. Further, there are enough tail events to help construct reasonable approximations of extremes. When it comes to events regarding human interactions, such as crime, illness, death or accidents, these are reported by statute to the relevant central authorities. This data is publicly available. In both these cases agents at all levels share the public signals and can condition their private expectations on good quality evidence. Of course, there may be variability in the accuracy of private expectations based on individual interpretation of the data or circumstances. This set-up allows the buyers of insurance the calculate their expected loss in a well informed manner and the insurance companies, based on the public information, can quote a premium. In turn the reinsurers share the same beliefs as no further information is available to them regarding the likelihood of the different states of nature. 

When it comes to cyber-risk and cyber-insurance, the state of data curation and sharing is far more nascent than for other insurance perils and it is reasonable to argue that there is no high quality public signal to inform all agents' priors. In the regulation of the aviation industry, it is standard to require reporting of `near misses' so that lessons can be learnt and procedures updated to lessen the risk of future accidents. It is possible that this might be addressed by vendor telemetry --- an insurer might have a series of recommended cyber-security solution providers that their clients could sign up for as part of their insurance package who would share data with the insurer. This raises potential issues of confidentiality. 

\subsection{Cyber-insurer loss experience}
\begin{figure}[!htb]
\centering
\includegraphics[width=12cm]{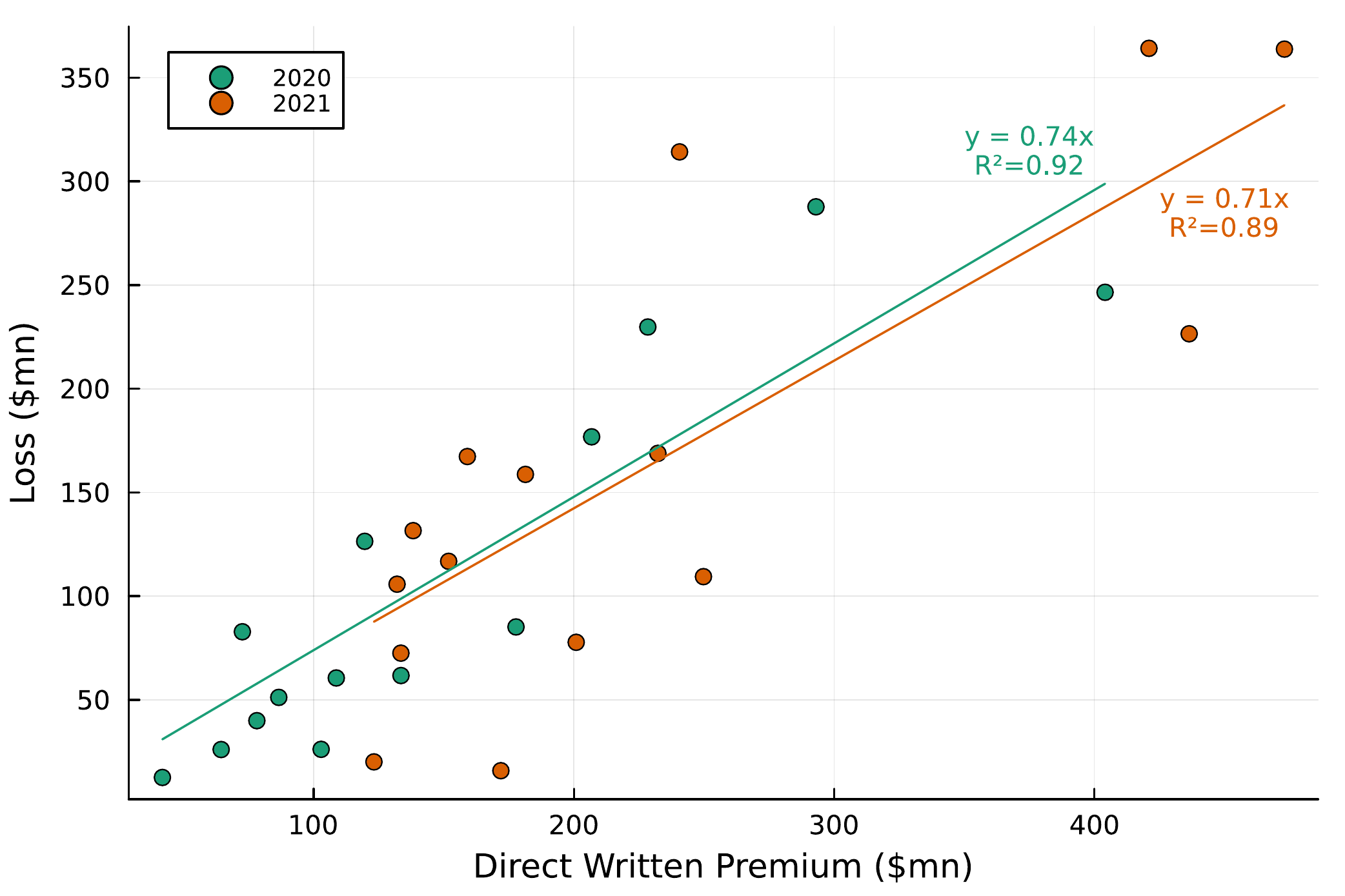}
\caption{US cyber-insurer losses vs premium written}
\label{fig:naic-cross-section}
\end{figure}

\begin{table}[!htb]
\begin{center}

\begin{tabular}{l|cccc|cccc}
     \hline
     &\multicolumn{4}{c|}{Direct Written Premium (\$mn)}&\multicolumn{4}{c}{Loss Ratio} \\
     \hline
     Firm&2018&2019&2020&2021&2018\textdagger&2019\textdagger&2020&2021\\
     \hline
     CHUBB LTD GRP&320.73&355.28&404.14&473.07&28.6\%&27.7\%&61.0\%&76.9\% \\ 
FAIRFAX FIN GRP&38.15&65.01&108.69&436.45&23.4\%&51.6\%&55.7\%&51.9\% \\ 
AXA INS GRP&255.87&229.68&293.03&421.01&57.2\%&65.7\%&98.2\%&86.5\% \\ 
TOKIO MARINE HOLDINGS&44.59&46.91&78.16&249.79&30.6\%&17.1\%&51.1\%&43.8\% \\ 
AMERICAN INTL GRP&232.31&226.20&228.42&240.61&36.1\%&55.4\%&100.6\%&130.6\% \\ 
TRAVELERS GRP&146.23&178.53&206.82&232.28&22.4\%&32.1\%&85.5\%&72.7\% \\ 
BEAZLEY INS CO INC&110.95&150.94&177.75&200.88&7.8\%&22.0\%&47.9\%&38.7\% \\ 
CNA INS GRP&83.36&94.72&119.61&181.38&26.9\%&33.2\%&105.7\%&87.5\% \\ 
ARCH INS GRP&---&---&---&171.94&---&---&---&9.2\% \\ 
AXIS CAPITAL GRP&76.00&97.31&133.55&159.06&7.2\%&18.5\%&46.2\%&105.2\% \\ 
ZURICH INS GRP&43.32&43.67&64.43&151.87&18.2\%&86.9\%&40.4\%&76.9\% \\ 
LIBERTY MUT GRP&66.50&68.38&41.86&138.22&38.9\%&23.3\%&30.0\%&95.2\% \\ 
SOMPO GRP&34.05&49.71&72.59&133.52&56.7\%&29.3\%&114.1\%&54.3\% \\ 
BCS INS GRP&69.50&76.06&86.58&132.04&10.4\%&32.9\%&59.1\%&80.1\% \\ 
HARTFORD FIRE 7 CAS GRP&39.70&49.74&102.86&123.16&16.4\%&31.6\%&25.4\%&16.3\% \\ 
     \hline
     
\end{tabular}
\vspace{1mm}
\caption{Cyber-insurer loss experience in the US market\\ (\textdagger   denotes weighted average by DWP)\\Source: NAIC, Researcher calculations}
\label{table:naic-figures}
\end{center}
\end{table}
The United States National Association of Insurance Commissioners publishes an annual report on the cyber-insurance market derived from its Property/Casuality Annual Statement~\cite{naic2021}. Table~\ref{table:naic-figures} presents this information for the four years currently available. In 2018 and 2019, the data was presented separately for standalone and package policies but in 2020 and 2021 was presented for combined policies. We have adjusted for this to present the data on a consistent basis. It is notable that the ransomware epidemic from 2020 to 2021 had a marked effect on experienced loss ratios for some insurers\footnote{This has been widely reported in the trade press -- see, for example, \cite{reutersransomware}}. However, there are pockets of differentiation. For example, the Hartford Insurance Company specialises in insurance for smaller companies, creating a fairly well diversified portfolio of insurance contracts where the holders are unlikely to fall victim to sophisticated, targeted ransomware attacks given the potential revenue available. For these companies, basic defences and security software should help mitigate against losses. Figure~\ref{fig:naic-cross-section} plots the losses experienced in the underwriting year versus the premium written and a linear trend line with intercept fixed at 0. The slope of the fitted trend line is then the loss ratio. The average loss ratio remained fairly stable across the two years, but it is striking that less than 30\% of premia received was, on average, retained by the underwriting insurer. The aforementioned NAIC report states that some 50\% of premia for cyber-insurance was ceded to the reinsurance market. There is some evidence to support the premise of a disconnect between expected and experienced losses in cyber-insurance pricing. Woods et al (2021)~\cite{woods2021county} develop a distribution of cyber-losses based on insurance company filings in the United States. They note that their model significantly under-predicts losses in relation to \textit{ex-post} losses reported in other literature. The under-pricing of premia implies that either

\begin{itemize}
    \item Insurers believe they can diversify loss risk.
    \item Customers were not willing to pay the fair premium and insurers are pursuing a `loss-leader' strategy. 
\end{itemize}

The entry of Arch Insurance also merits comment. Arch insurance provides capacity\footnote{\url{https://www.coalitioninc.com/en-ca/announcements/Arch-Insurance-Backs-Coalition-With-Long-term-Capacity-Across-Cyber-Insurance-Programs}} to a relatively new managing general agent (MGA), Coalition Inc., providing `active cyber-insurance'. Active cyber-insurance is a relatively new product, which merges the roles of an outsourced security provider and a traditional cyber-insurer. This reduces some of the risks of asymmetric information transfer associated with cyber-insurance from the perspective of the insurer. The trade-off between cyber-insurance and security investment has been modelled by Mazzoccoli and Naldi (2020)~\cite{mazzoccoli2020robustness} and Skeoch (2022)~\cite{skeoch2022expanding}.

Moves to limit potential catastrophic cyber-insurance losses have been undertaken by Lloyd's of London, one of the world's leading insurance markets. In August 2022, Lloyd's reportedly announced\footnote{See, for example, \url{https://www.insurancetimes.co.uk/news/lloyds-of-london-sets-cyber-policy-exclusions-for-state-backed-attacks/1442106.article}} that insurers should not provide insurance coverage for state-sponsored attacks. This in theory reduces the risk of a castrophic cyber-event as the type of vulnerabilities exploited in such an attack and the sophistication and scale of resources needed to exploit them are usually only found in states with significant offensive cyber capabilities. 

\subsection{Loss transparency}
We consider what happens if agents only selectively claim on losses from an insurer. In an insurance analysis, it is usually assumed that every agent is aware of the attacks they experience. This is a reasonable assumption for some categories of cyber-attacks, such as ransomware, although other cyber-attacks such as data breaches might not be detected until some time after the event. Agents report some attacks to an insurer and thus a claim is made; some attacks go undisclosed (in insurance, this is known as IBNR --- incurred but not reported). More formally, at time $t$, the agent may be aware of the attack and its damage so the state of the world in which the attack occurs, $s$ is known to them. The agent might inform the insurer about the state so the insurance knowledge of the state $s$ is conditional on the revelation of the agent. Now, the insurer knows that their distribution is not the objective one but only a partial revelation due to the agents selectively choosing to report losses. The insurer then tries to approximate the objective distribution but it will be with error. In the event that reinsurers know that different insurers have different approximations of the true distribution, they will use some kind of averaging across these approximations to quote reinsurance premiums. The results are: 
\begin{itemize}
    \item No insurer is offered a fair premium given their approximation of the true distribution.
    \item No agent is offered a fair premium as the insurance offer is based on a distribution different to their own.
    \item Objectively measured data  is absent at all levels because reporting is a choice.
\end{itemize} 

\subsection{Consistency of reference}

There is a significant problem with the standard actuarial modelling cycle approach to cyber-insurance: the evolution of systems over time, which is quite unique in its complexity in relation to other perils. Calibration of models using events such as WannaCry have poor future predictive power as the security vulnerabilities it exploited have been patched, Windows XP is less widespread than it was and the operating systems that replaced it have better, though of course not perfect, security by design. In economics, this can be couched in clients' Bayesian updating of their distributions; they do not and cannot observe attacks on other clients (other than indirectly via media reports) so there is no need to converge to a stationary distribution at the client level. The consequence of this is that the insurers and reinsurers may have a better understanding of the fair price of risk, but buyers do not share the same concern and thus are not willing to pay the demanded premium for the insurance. 

\subsection{Supply and demand}

In the insurance industry, it is common to describe the state of the market as `hard' or `soft'. In a soft market, supply exceeds demand placing downward pressure on premium, whereas in a hard market the converse is true. Often the experience of losses in a particular class of business will result in a market hardening. This has important implications for the pricing of cyber-insurance by a vendor. In a soft market, the insurer must charge the lowest premium it can actuarially justify to build market share. In a hard market, the insurer should charge the highest realistic premium possible. If the market were efficient, it would converge to some form of equilibrium but if not it may swing between financial imbalances. There is evidence that in the early stages of the cyber-insurance industry, some insurers operated a very experimental approach to pricing. Woods (2023)~\cite{woods2023turning} provides an account of one large US insurer, whose Chief Operating Officer admitted that their early cyber-insurance models were a ``complete guess''. The same insurer then suffered loss ratios of 100\% and 130\% in 2020 and 2021, respectively (Table~\ref{table:naic-figures}), suggesting that even if refined and updated, the pricing models may have underestimated the claim frequency or severity.

\subsection{Further Work}

We have considered simulations in which losses are uncorrelated. An interesting next step would be to consider the correlation of losses and implement the modeling strategy presented in this paper using more complex loss-generating functions, such as those reviewed in Section~\ref{subsec:actuarial-models}, than the simple joint distributions of severity and frequency used in this paper. It would also be instructive to compare the results of simulations of distributions proposed by Eling et al (2019)~\cite{eling2019actual} and Woods et al (2021)~\cite{woods2021county}, with insurer loss data. Claims data is deeply confidential to insurance companies, however, so the results of such analysis would unlikely be able to be widely disseminated unless extensively anonymised. 

In the simulations, we focused on the supply dynamics of insurance and in particular the interaction between insurers and reinsurers. The model provides for consideration of buyer preferences, which at this stage we have explored only briefly in the first simulation to illustrate how buyer utility can affect coverage. A further piece of work would be to explore the price sensitivity of buyers of insurance coverage and how these preferences propagate through the information chain to reinsurers. 

\subsection{Conclusions}

This paper has developed an artificial yet realistically structured model of the cyber-insurance market considering all three levels of agent interactions.  The model incorporates the demand choices of the consumers/buyers of cyber-insurance, their suppliers --- insurance companies offering contracts --- and reinsurance companies providing additional underwriting capacity. 

The extent to which an insurance market facilitates smooth risk transfer is linked to the sharing of information by participants regarding the distribution of losses. We argue that this condition is very unlikely to hold in the cyber-insurance market. Disagreements on loss expectations means that cyber-insurance contact pricing will be considered inefficient at both the retail and wholesale levels, leading to lower societal benefit. The purpose of this paper was to quantify such inefficiency within the confines of a three-tier market under miscellaneous types of disagreements in loss expectations among the participants at each tier. 

To establish a benchmark to gauge the extent of inefficiency, we have simulated a simple market where all agents share a distribution of losses based on two loss frequencies. From this simulation, we obtained the ‘efficient’ measures of reinsurance premium and the proportional participation of reinsurers. We found that simulated loss reduction to the insurers is almost identical to the cost of reinsurance (bar small statistical errors), as expected. This case represents the economically efficient market outcome. 

Maintaining all the behavioural parameters from the first simulation, we then proceeded to compute expected losses and reinsurance premiums based on diverse distributions held by insurance companies and reinsurers. Both insurers and reinsurers independently price premiums to meet target loss ratios based on distinct and subjective distributions. Under conditions where losses are close to the modal simulated value, insurers are typically not incentivised to buy reinsurance. However, when considering relatively extreme losses under a `stress test' type scenario, the value of reinsurance emerges to some insurers whose distributions are relatively heavy tailed in comparison to others. For such insurers, the upfront cost of such reinsurance is justified by the avoidance of ruin under high loss scenarios.   

Even within the confines of this simple example, the divergence in distributions, expectations and objectives demonstrates that efficient pricing is hard to achieve. It should be noted that whilst there are specialists in cyber-insurance operating within the reinsurance market, cyber-insurance itself competes with other lines of insurance for allocation of Specialty reinsurance capital. Based on this, we used a uniform cost of reinsurance in the second of our two simulations. This is the outcome of the reinsurer holding a private loss distribution. This condition may reduce the reinsurance capital allocated to cyber-insurance.  

  Our findings suggest that the cyber-insurance market will continue to face potential financial imbalances. That is, it will be highly profitable for some participants and costly for others. This is already evident in data on cyber-insurer loss data (Table~\ref{table:naic-figures}). There has been considerable progress in the academic literature on theoretical modelling of cyber-losses and on empirical analysis. However, access to reliable and transparent data remains a problem for researchers as insurance claims data is confidential and highly guarded. Braun et al (2023)~\cite{braun2023cyber} have noted that an insurance-linked securities market to support cyber-insurance may struggle to develop without better cyber-modelling. Without a means of accessing reliable data on cyber-losses, insurance buyers will have to continue to form highly subjective probability. The absence of statutory reporting of cyber incidents is a major issue that needs to be addressed. In a recent paper, Bajoori et al (2022)~\cite{bajoori2022} argue for the creation of an official registry of cyber-security experts with a duty to report. This has also been proposed by the UK Government\footnote{https://www.ncsc.gov.uk/information/ncsc-assured-cyber-security-consultancy}. 

The cyber-insurance market is still at as stage of relative infancy. The current institutional setup does not appear fully conducive to the delivery of efficient market outcomes at this juncture. Achieving efficiency requires commonly held beliefs and stationary loss distributions. Whether such conditions can be achieved and maintained is questionable given the dynamic nature of cyber-threats. Our provisional conclusions are that the most likely market structure will involve firms specialising in particular insurance contracts covering different ranges of loss limits, with varying access to reinsurance based on these contracts. The overall outcome will be that the capital capacity of this market will be below its optimal size under shared informational conditions.  

\section*{Acknowledgements}

Henry Skeoch was employed by Convex Insurance between July-August 2021, which provided the inspiration for part of this work. He would like to thank Christophe Chandler, Rob Smart and Harry Thompson in particular for helpful discussions during this time and for sharing their knowledge and acumen on the reinsurance markets. The arguments and analysis presented in this research are solely those of the authors and should not be interpreted as representative of the views or business practices of Convex Insurance, its clients or its counterparties. HS and CI would like to extend particular thanks to Martin Eling for most helpful comments and recommendations that have greatly improved the structure and substance of the arguments within this paper. HS would like to thank Marie Vasek for helpful discussions as well as his fellow PhD candidates at UCL for their support and encouragement throughout this work. Anonymous reviewers for the Workshop on the Economics of Information Security 2023 provided helpful feedback and suggestions. This work was supported partly by the UK Engineering and Physical Sciences Research Council grant for Doctoral Training EP/R513143/1. 

\clearpage

\printbibliography

\appendix
\section{Insurer Loss Distributions} \label{appendix}
\begin{figure}[!htb]
\centering
\includegraphics[width=16cm]{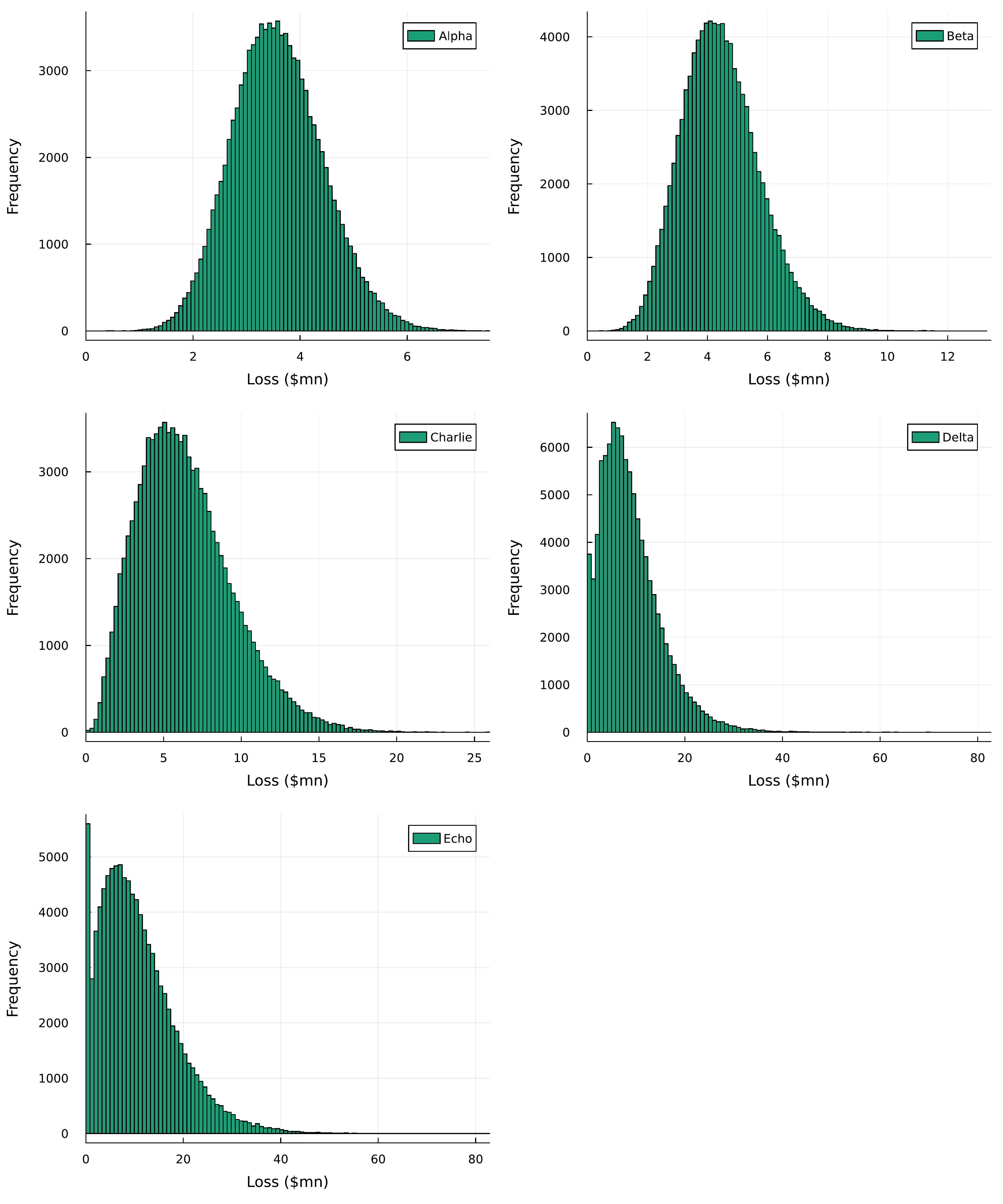}
\caption{Insurer Simulated Loss Distributions (Section~\ref{subsubsec:simul-proc})}
\label{fig:appendix-sim-loss-dist}
\end{figure}

\end{document}